\documentclass[5p,twocolumn]{elsarticle}

\usepackage{hyperref}
\usepackage{svg}
\usepackage{graphicx}

\usepackage{tabularray}

\usepackage{caption}
\usepackage{subcaption}
\usepackage{adjustbox}
\usepackage{tcolorbox}
\usepackage{color}

\usepackage{multirow} 
\usepackage{threeparttable}

\usepackage{array} 
\usepackage{makecell} 
\usepackage{booktabs} 

\usepackage{booktabs}
\usepackage{xcolor}
\usepackage{colortbl}

\usepackage{url}

\usepackage[switch, modulo]{lineno}
\begin{document}
\begin{frontmatter}

\title{Insights into resource utilization of code small language models serving with runtime engines and execution providers}

\author[upc]{Francisco Durán\corref{cor1}}
\ead{francisco.duran.lopez@estudiantat.upc.edu}

\author[upc]{Matias Martinez}
\ead{matias.martinez@upc.edu}

\author[vu]{Patricia Lago}
\ead{p.lago@vu.nl}

\author[upc]{Silverio Martínez-Fernández\corref{cor1}}
\ead{silverio.martinez@upc.edu}

\cortext[cor1]{Corresponding author}

\affiliation[upc]{organization={Universitat Politècnica de Catalunya},
            city={Barcelona},
            state={Catalunya},
            country={Spain}}

\affiliation[vu]{organization={Vrije Universiteit Amsterdam},
            city={Amsterdam},
            country={Netherlands}}





\date{July 2025}

\begin{abstract}

The rapid growth of language models, particularly in code generation, requires substantial computational resources, raising concerns about energy consumption and environmental impact. Optimizing language models inference resource utilization is crucial, and Small Language Models (SLMs) offer a promising solution to reduce resource demands.
Our goal is to analyze the impact of deep learning serving configurations, defined as combinations of runtime engines and execution providers, on resource utilization, in terms of energy consumption, execution time, and computing-resource utilization from the point of view of software engineers conducting inference in the context of code generation SLMs.
We conducted a technology-oriented, multi-stage experimental pipeline using twelve code generation SLMs to investigate energy consumption, execution time, and computing-resource utilization across the configurations.
Significant differences emerged across configurations. CUDA execution provider configurations outperformed CPU execution provider configurations in both energy consumption and execution time. Among the configurations, TORCH paired with CUDA demonstrated the greatest energy efficiency, achieving energy savings from 37.99\% up to 89.16\% compared to other serving configurations. Similarly, optimized runtime engines like ONNX with the CPU execution provider achieved from 8.98\% up to 72.04\% energy savings within CPU-based configurations. Also, TORCH paired with CUDA exhibited efficient computing-resource utilization.
Serving configuration choice significantly impacts resource utilization. While further research is needed, we recommend the above configurations best suited to software engineers’ requirements for enhancing serving  resource utilization efficiency. 

\end{abstract}


\end{frontmatter}

\section{Introduction}\label{sec:introduction}

The widespread adoption of deep learning models, particularly language models, has surged remarkably in recent years. This trend has led to an exponential increase in the computational resources needed to support these models~\cite{shoeybi2019megatron}, raising concerns such as environmental impacts \cite{Strubell2019Monitoring}. The focus within the deep learning community has predominantly been on enhancing model accuracy, often overlooking the associated costs. For instance, Benchmarks like GLUE or SuperGLUE encourage competition based on accuracy, often without much consideration for reducing energy consumption \cite{cowls2023ai}. State-of-the-art models like GPT-4 and Claude 2 demand significant computational resources, primarily accessible to companies like OpenAI, Anthropic, or Google, leading to a large environmental footprint \cite{perrault2024artificial}. Notably, Schwartz \emph{et al.} \cite{Schwartz2020GreenAI} emphasized the need for deep learning research to prioritize efficiency, advocating for a balance where improvements in accuracy are not achieved at the expense of a larger carbon footprint. 

The escalating computational demand of deep learning models, has led to an increase in research focused on improving the sustainability of these systems, including studies that evaluate deep learning architectures such as BERT, DenseNet, and RNNs in NLP and computer vision contexts \cite{dodge2022measuring, ajel2022energy}. Although the majority of this research targets the training phase, a growing body of work acknowledges the significance of both training and inference in achieving a greener deep learning \cite{ martinez2023towards, bhatt2024towards}. However, there remains a noticeable gap in studies dedicated to the inference phase \cite{verdecchia2023systematic}. This oversight is important as inference can also contribute substantially to the overall energy footprint of deep learning systems, especially when deployed at scale \cite{zhao2022green}. Addressing this gap is crucial to reduce overall environmental impact and encourage the development of resource-efficient deep learning systems.

During inference, trained models are integrated into deep learning systems to allow users to make predictions, a process known as deep learning serving \cite{zhang2020inferbench, kumara2023requirements}. Critical and popular architectural decisions in deep learning serving include the selection of serving infrastructure, with the runtime engine being one of the most commonly used options, and the choice of an execution provider.
These decisions are driven by the community's focus on optimizing performance and simplifying deployment to handle growing deep learning systems' complexity \cite{huggingfaceModelDeployment, huggingfaceTools}.
The runtime engine is a software component that loads the model, performs inference, and returns results to the user. It can also apply optimizations, such as graph optimizations or just-in-time compilation \cite{duran2024identifying}. The execution provider, which is the library that serves as the backend of the runtime engine, optimizes model inference based on the available hardware \cite{onnxEP, alizadeh2024green}. These two key decisions are able to significantly affect the 
resource utilization\footnote{\url{https://din.one/site/rs}}.
In line with the ISO 25010~\cite{iso25010}, resource utilization is identified as a key subcharacteristic of the broader quality attribute of efficiency. This attribute refers to how effectively a system uses resources under stated conditions.



In this paper, we investigate the 
resource utilization in terms of energy consumption, execution time, and computing-resource utilization of different configurations of deep learning serving infrastructures, focusing on runtime engines and execution providers.  We concentrate on one particular deep learning task, specifically code generation, as it is one of the most popular tasks of language models and has a potential impact on software development productivity \cite{ziegler2022productivity, gu2025effectiveness}. As these models are increasingly integrated into development environments, optimizing their inference efficiency becomes highly relevant for practitioners concerned with both performance and sustainability.
Code generation language models have demonstrated impressive capabilities in code generation \cite{xu2022systematic}. Their success largely stems from the Transformer architecture, which excels at learning complex patterns and scales efficiently \cite{vaswani2017attention,jiang2023impact}. 
This has enabled state-of-the-art performance in models like Codex \cite{chen2021evaluating}, the backbone of GitHub Copilot\footnote{\url{https://copilot.github.com/}}, a tool that provides intelligent code suggestions to developers.

These advanced code generation language models, however, often contain millions to billions of parameters \cite{xu2022systematic, zan2023large}, requiring vast amounts of training data and substantial computational resources. In contrast, Small Language Models (SLMs) have emerged as a promising alternative, offering similar capabilities but with fewer computational demands, making them accessible to individual developers \cite{wang2024comprehensive}.
SLMs refer to language models with parameter counts typically below the scale of tens of billions \cite{zhu2024llava, hu2024minicpm}, a threshold that continues to evolve as the field advances \cite{wang2024comprehensive, van2024survey}. This distinguishes them from Large Language Models (LLMs), which often exceed tens or hundreds of billions of parameters. Code generation SLMs refer specifically to a subset of small language models trained on source code and capable of demonstrating strong performance in code generation tasks, either as their primary application or as a key evaluation benchmark. 
Given their efficiency and lower resource requirements, our research focuses specifically on code generation SLMs, a field where further exploration could lead to significant advancements in research and software development \cite{perrault2024artificial}. 
Specifically, we focus on the auto-complete code task, in which a SLM continuously predicts the next tokens as the developer types. This is a standard feature in modern integrated development environments (IDEs), designed to boost productivity and minimize keystrokes by generating a few tens of tokens from short input prompts \cite{weber2024significant, jiang2024survey}. 

We evaluated the impact of various deep learning serving decisions, specifically the choice of runtime engine and execution provider, through a technology-oriented experiment using twelve code generation SLMs \cite{wohlin2012experimentation}. The experiment utilized a generated input dataset from the HumanEval benchmark \cite{chen2021evaluating}, allowing us to evaluate the energy consumption, execution time, and computing-resource utilization of each serving configuration. These configurations were based on combinations of different runtime engines and execution providers. We tested several runtime engines, including  Torch (TORCH), ONNX Runtime (ONNX), OpenVINO Runtime (OV), and Torch JIT (JIT), across both CPU and GPU execution providers, with the aim of identifying the most efficient serving configurations.

Our results revealed significant differences in energy consumption, execution times, and computing-resource utilization across the various serving configurations. Configurations that employed the CUDA execution provider consistently outperformed those using the CPU execution provider in both energy and time efficiency. The combination of TORCH with the CUDA execution provider, as well as ONNX or OV with the CPU execution provider, yielded the best results for those metrics. JIT can perform effectively in certain deep learning tasks, such as computer vision \cite{goldfarb2024evaluating}, but it needs further refinement to handle language generation scenarios such as code generation, as discussed later in this paper. Overall, the TORCH engine paired with the CUDA execution provider demonstrated efficient utilization of computing resources.

The contributions of this work are:
\begin{itemize}
\item \textbf{Actionable guidelines for software engineers:} 
We draw practical guidelines for software engineers under varying serving infrastructure constraints. While these are derived from our experimental context, they are intended to support broader serving decisions in industry and academia. The guidelines address common serving trade-offs, particularly relevant to environments where sustainability, latency, or computing-resource constraints are key. These will be elaborated in detail in Section \ref{sec:discussions}.
\item \textbf{Measuring the impact of deep learning serving configurations on resource utilization:} CUDA execution provider configurations consistently reduce energy consumption and execution time during inference. Using TORCH as a runtime engine and CUDA execution provider offers both optimal energy efficiency and faster execution times, with energy savings ranging from 37.99\% to 89.16\% and execution time reductions from 47.84\% to 89.74\% compared to the worst-performing runtime engine (ONNX Runtime) in CUDA execution provider configurations. The improvements are greater when evaluated against its CPU counterparts.
The study emphasizes the importance of monitoring resource usage to prevent bottlenecks, such as high CPU or RAM usage in CPU execution provider configurations. It highlights the trade-offs of using optimized runtime engines, balancing energy and execution time improvements against increased resource demands, and showcases the consistent resource efficiency of the baseline runtime engine, TORCH, across various execution providers.
\item \textbf{A reproducible pipeline and analysis of deep learning serving configurations:} We developed a multi-stage experimental pipeline to evaluate various deep learning serving infrastructures. This pipeline tested the resource utilization across different configurations.

\end{itemize}

\textbf{Data availability statement}. The replication package is publicly accessible\footnote{Please refer to \url{https://doi.org/10.5281/zenodo.15258679}} following the ``Cookiecutter Data Science'' project structure\footnote{\url{https://cookiecutter-data-science.drivendata.org/}}. It includes the extracted model metadata, the developed serving infrastructure, profiling data, data analysis, statistical tests results, results for each model, and comprehensive documentation to support reproducibility and further research.

The paper is organized as follows. In Section~\ref{sec:background}, we introduce the background of our research. In Section~\ref{sec:related_work}, we introduce the pertinent prior research. Moving to Section~\ref{sec:research_methodology}, we present the research questions and the study design. Subsequently, Section~\ref{sec:results} presents the findings. Section~\ref{sec:discussions} shows the discussions and implications. The limitations and threats to validity are described in Section~\ref{sec:threats}. Finally, Section~\ref{sec:conclusions} draws conclusions and outlines directions for future work.

\section{Background}\label{sec:background}

In this section, we introduce (i) the deep learning serving configurations: duplet of a runtime engine and an execution provider; (ii) the type of deep learning models we use, namely code generation SLMs.

\subsection{Deep learning serving configurations}

\begin{figure*}[htp]
    \centering\includegraphics[width=\textwidth]{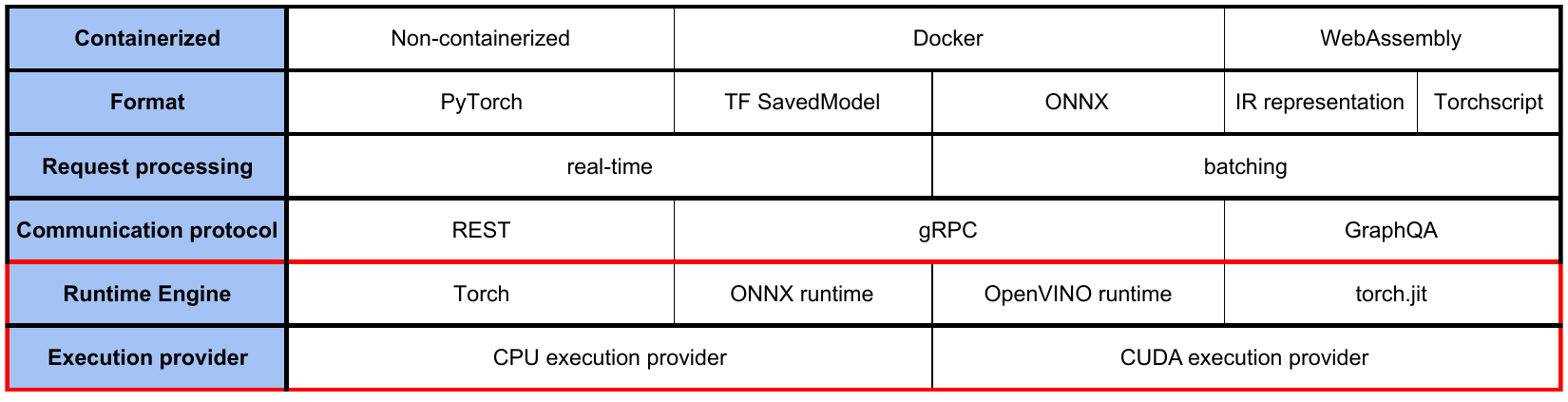}
    \caption{Architectural design decisions for deep learning serving: runtime engine as a serving infrastructure and execution provider options highlighted in red. Figure adapted from \cite{nvidiaEP}.}
\label{fig:adds_re_ep}
\end{figure*}

In previous work \cite{duran2024identifying}, we identified four distinct types of deep learning serving infrastructures within the deep learning serving architectural design decisions: i) No runtime engine and web server, ii) Runtime engine and web server, iii) deep learning-specific software, and iv) End-to-end deep learning cloud service. Figure~\ref{fig:adds_re_ep} illustrates the first two serving infrastructures alongside their transversal decisions. While these transversal decisions can improve the interaction between users and the deep learning application, they are influenced by the underlying serving infrastructure. Additionally, we include the execution provider, which acts as the backend of the runtime engine, optimizing execution for specific hardware.

\textbf{Runtime engine.} The runtime engine is a software component that loads the model into the system's memory from a model registry, performs inference, and returns the results to the user. Optimized runtime engines (e.g., ONNX runtime, OpenVINO runtime, or Torch JIT) offer a flexible interface that facilitates the integration of hardware-specific libraries and implements model optimizations. 
These include graph optimizations like constant folding and elimination of redundant nodes, as well as just in time compilation \cite{duran2024identifying, onnxRE, onnxruntime2024cuda}. 
This setup can operate independently or be integrated with a web server, depending on the serving requirements. For instance, the optimized ONNX runtime can be used to load ONNX models and wrap them as a service using a web framework like FastAPI. Once the service is running, the user can make predictions through the service, using a communication protocol like gRPC (see Figure 6 from \cite{zhang2020inferbench}). 
Additionally, there is the option of using the deep learning framework directly without an optimized runtime engine. This is the simplest approach, as deep learning frameworks are primarily designed for training but can also be used for inference \cite{georgiou2022green}. Several runtime engines are available, including:

\begin{itemize}
    \item \emph{Torch (TORCH)}. Inference is conducted directly through the deep learning framework, for instance, the PyTorch library. This approach relies on the framework's inherent capabilities for model execution without additional runtime optimizations. This approach serves as the baseline, representing a straightforward and unoptimized runtime engine.

    \item \emph{ONNX Runtime (ONNX)}. The inference is made through a runtime engine, which applies optimizations to the ONNX model. This is one of the most widely used options, with tools like the Optimum library facilitating the use of the ONNX runtime. It is the default optimized runtime for models in ONNX format \cite{zhang2020inferbench}.

    \item \emph{OpenVINO Runtime (OV)}. The inference is made through the OpenVINO Toolkit \cite{intel2019openvino} runtime engine, which applies optimizations through an Intermediate Representation (IR) and is converted using its Model Optimizer. This runtime engine has been the subject of several studies for its efficiency \cite{lenherr2021new}, and libraries like Optimum provide support for OpenVINO runtime integration \cite{intel2019openvino}.

    \item \emph{Torch JIT (JIT)}. A runtime engine from PyTorch, it compiles PyTorch models into TorchScript, an intermediate representation that can be run independently of Python. It is the default optimized runtime for models in TorchScript format \cite{zhang2020inferbench}.
    
\end{itemize}

\textbf{Execution providers.} These are libraries that act as the back-end of the runtime engine, optimizing execution for specific hardware (CPU, GPU, FPGA, etc.) \cite{alizadeh2024green, onnxEP}. 
They perform kernel level optimizations: selecting, fusing or tuning the fine grained operations, called kernels, that span from an individual convolution to a fused block combining convolution, activation and normalization \cite{nvidia2021gtc,nvidia2022onnxblog,meng2020differentiable}. The specific kernel variant is chosen when the model is loaded, either through lightweight heuristics or a full exhaustive search, so that inference runs on the fastest implementation available for the detected hardware \cite{nvidia2022onnxblog}.

There are various execution providers, including:

\begin{itemize}
    \item \emph{CPU Execution Provider (CPU)}. Default backend library for CPU \cite{onnxEP}.

    \item \emph{CUDA Execution Provider (CUDA)}. CUDA is the most popular backend library and enables hardware accelerated computation on NVIDIA CUDA-enabled GPUs \cite{alizadeh2024green, nvidiaEP,nvidiacuda}.
\end{itemize}

\subsection{Code generation Small Language Models (SLMs)}

Increasing evidence indicates that SLMs deliver remarkable results in code-related tasks, such as code generation, achieving comparable accuracy and output quality to LLMs despite having significantly fewer parameters \cite{hsieh2023distilling, gunasekar2023textbooks, schick2020s}. This is pertinent for practitioners using limited resources or concerned with privacy issues \cite{shi2024greening, svyatkovskiy2021fast}, situations where using LLMs is neither feasible nor desirable. 

In this work, we define code generation Small Language Models (SLMs) as language models that have fewer than 5 billion parameters specialized in generating source code. Our definition aligns with the perspective presented in \cite{wang2024comprehensive}, where, given specific tasks and resource constraints, SLMs fall into a range bounded by a minimum size at which the model first exhibits emergent capabilities for a specialized task, and an upper bound dictated by available computational resources. Although the exact scale of SLMs may evolve as new models emerge \cite{van2024survey}, their core objective remains stable: achieving effective performance within constrained computational conditions. Thus, in our specific context, code generation task, SLMs are models that demonstrate proficient capabilities comparable to larger models in the code generation task, but under significantly restricted computational and resource constraints compared to state-of-the-art LLMs.

Among the most prevalent real-world SLMs applications, we concentrate on the auto-complete code task of code generation, where the model observes partially written code and predicts the next few tokens, typically on the order of a few tens of tokens. This setting replicates the interactive completion facilities embedded in modern IDEs, which raise developer productivity by providing token level suggestions at the moment of typing \cite{weber2024significant}. Experience reported by practitioners further shows that conversational code assistants often deliver larger fragments, whereas auto-complete interfaces prefer more frequent but smaller suggestions. These concise few-tokens completions are easier to inspect and validate while developing, while larger blocks demand greater cognitive effort before they can be safely incorporated \cite{weber2024significant, kalliamvakou2022research}.

Moreover, relying on closed-source LLMs like OpenAI's ChatGPT accessible via API can introduce prohibitive costs, privacy concerns, and disproportionate energy consumption for single inferences for few tokens \cite{khowaja2024chatgpt}. The exclusive capability of big tech companies to deploy state-of-the-art LLMs, due to their extensive computational and financial resources, leaves small to medium tech companies facing a significant challenge \cite{zhou2023opportunities}. These smaller companies often operate with constrained CPU or GPU resources, necessitating innovative approaches to make use of language models effectively. In this context, SLMs emerge as a critical solution, not only enabling a wider range of companies and deep learning practitioners to harness the power of deep learning but also contributing to environmental sustainability efforts by reducing the carbon footprint associated with large-scale computations \cite{zhou2023opportunities}.

\section{Related work}\label{sec:related_work}
There is a notable gap in studies addressing the deployment efficiency of trained deep learning models, despite well-established evidence that inference costs can exceed training costs in deployed systems due to the cumulative effect of using the model or system many times. Reports from major technology companies such as Amazon and NVIDIA estimate that inference may account for up to 90\% of total costs in production \cite{desislavov2023trends, perrault2024artificial}.

Studies like Zhang \emph{et al.} \cite{zhang2018} examined hardware and software impacts on inference efficiency, while Georgiou \emph{et al.} \cite{georgiou2022green} compared TensorFlow and PyTorch's energy and performance. Koubaa \emph{et al.} \cite{koubaa2021cloud} explored deployment strategies and model formats. Other works, including Hampau \emph{et al.} \cite{hampau2022empirical} and Klimiato \emph{et al.} \cite{klimiato2022utilizing}, investigated containerization and deep learning serving infrastructures' effects on energy and performance. Escribano \emph{et al.} \cite{escribano2023energy} focused on cloud computing energy consumption for NLP models, omitting deep learning serving infrastructure comparison.

A recent exploration of the code generation language models ecosystem by Yang \emph{et al.}  \cite{yang2024ecosystem} offers a comprehensive analysis of code generation language models, shedding light on ecosystem dynamics, model reuse strategies, and documentation practices. Their findings emphasize emerging approaches such as quantization and fine-tuning, which could directly impact deployment efficiency and robustness. 

Recent studies have examined language models for the code auto-complete task, highlighting their impact on developer productivity and system performance \cite{weber2024significant}. GitHub reports up to a 55\% speed boost when using their tool GitHub Copilot in completion mode \cite{kalliamvakou2022research}. There is a growing demand to deploy these models within IDEs on local machines. However, even moderately sized models can introduce significant inference overheads like higher latency, increased energy use, and a larger carbon footprint \cite{shi2024greening}. To address these concerns, SLMs have been proposed for IDE integration \cite{svyatkovskiy2021fast}, and dynamic inference methods, such as GREEN-CODE's energy-aware early exits \cite{ilager2025green} and GPT-2's first-layer predictions \cite{sun2024neural}, demonstrate substantial efficiency gains when generating only a few tokens from short prompts. Yet, few works examine resource utilization during serving in the short prompt context typical of real-world auto-complete scenarios.

Few studies have considered the runtime engine in the inference phase. 
Hampau \emph{et al.} \cite{hampau2022empirical} analyzed the resource utilization of three different containerization strategies. Their study focused on computer vision tasks and evaluated the following strategies: a single runtime engine without containerization, the same runtime engine with Docker, and WebAssembly (WASM). In contrast, our work addresses two distinct dimensions of the deep learning serving infrastructure: the choice of runtime engine and execution provider.
Zhang \emph{et al.} \cite{zhang2020inferbench} conducted a large study on four different deep learning Serving Infrastructures: two deep learning-specific software and two runtime engines. Although they explored the performance of ONNX Runtime and Torch.JIT, their findings primarily reported on latency, with experiments conducted on versions containerized using Docker and communicated through gRPC API (see Figure~\ref{fig:adds_re_ep} for a reference of the different design decisions).
With regard to execution providers, Alizadeh \emph{et al.} \cite{alizadeh2024green} analyzed the performance and energy differences between CUDA and TensorRT when paired with the ONNX runtime engine and a ResNet model, finding that TensorRT outperformed CUDA. In our research, we examine the default CPU execution provider and CUDA execution provider across multiple code generation SLMs.

The field of software architecture for deep learning based systems is still evolving, presenting significant challenges for practitioners in making design decisions that optimize efficiency \cite{warnett2022architectural, Schwartz2020GreenAI}. While some studies have proposed architecture-based approaches to improve the efficiency and sustainability of deep learning systems \cite{bhatt2024towards, jepsen2024architecting}, research specifically addressing the serving infrastructure of these systems remains limited \cite{duran2024identifying}. 
This work contributes to the field by addressing the underexplored area of serving infrastructure selection, complementing existing efforts in architectural decision-making for greener deep learning systems.

Contrary to previous research, our study focuses on analyzing the resource utilization of various deep learning serving configurations during the inference phase of SLMs for code generation. Our analysis specifically targets runtime engines and execution providers, and their impact on code generation SLMs, evaluating critical factors such as energy consumption, execution time, and computing-resource utilization.

\section{Research methodology} \label{sec:research_methodology}

\subsection{Research goal and questions}

The research goal \cite{basili1994goal} of this study is to analyze \textit{deep learning runtime engines and execution providers} for the purpose of \textit{understanding their impact} with respect to \textit{resource utilization in terms of energy consumption, execution time, and computing-resource utilization} from the point of view of \textit{software engineers conducting inference} in the context of \textit{code generation SLMs}. 

Therefore, we intend to address the following research questions (RQ):

\begin{itemize}

\item [RQ1:] How is the energy consumption affected by the execution provider and runtime engine used for inference from code generation SLMs?

\item [RQ2:] How is the execution time affected by the execution provider and runtime engine used for inference from code generation SLMs?

\item [RQ3:] How is the computing-resource utilization (CPU usage, RAM usage, GPU usage, GPU memory utilization, and GPU used memory) affected by the execution provider and runtime engine used for inference from code generation SLMs? 

\end{itemize}

\subsection{Experimental units} \label{experimental_units}

In this subsection, we detail the experimental units and the process used to construct the input dataset for our study. The experimental units refer to the selected code generation SLMs used for inference, while the dataset construction draws from the widely-used HumanEval dataset \cite{chen2021evaluating}. The experimental units were subjected to treatments in the form of different deep learning serving configurations to evaluate them during inference.

The selection process for the code generation SLMs in this study was conducted with a systematic and structured approach \cite{castano2024lessons}, reflecting the state of available models as of November 6th, 2024. Given the time required for executing experiments, and analyzing results, this study might not include other recent code generation SLMs after such date. 

Using the Hugging Face API, metadata for 104,816 text-generation decoder models was retrieved, including detailed model card information. A series of filters were applied to refine the selection: models exceeding 5 billion parameters were excluded to focus on SLMs, and priority was given to models with potential HumanEval results explicitly mentioned in their metadata, model cards, or flagged as gated models. Only base models, identified within their respective model trees, were retained, narrowing the pool to 340 models potential candidates. 

Afterwards, through metadata inspection and structured filtering based on the explicit mention and reporting of HumanEval benchmarks \cite{chen2021evaluating}, we narrowed down the selection to 29 pretrained base models. `potential HumanEval results' refers specifically to models whose metadata explicitly indicated having undergone HumanEval evaluation, whether in structured through the HuggingFace API or textual forms. We manually verified the model metadata to ensure that the selected models are base pretrained models, not subjected to quantization or fine-tuning.

Given practical constraints, such as total experimental runtime and available infrastructure resources, we estimated experiment durations and concluded that exhaustively testing all 29 models across all configurations was infeasible. Consequently, we randomly selected 12 models, presented in Table~\ref{tab:models_table}, intentionally ensuring diversity and representativeness of different sizes (e.g., 110 [M], 1 [B], and 3[B]), affiliation (e.g., academia and industry), and release dates (from 2021 to 2024). A replication package containing the extracted data is available for reproducibility and future research (see data availability statement in Section~\ref{sec:introduction}).

Our primary focus remained on understanding the impact of the serving infrastructure rather than extensively characterizing individual models. Nevertheless, we confirmed through preliminary tests that model outputs remained consistent across configurations, except for JIT configurations, whose inconsistencies were explicitly investigated and documented within our publicly available replication package.

\begin{table*}[t]
    \centering
        \caption{Overview of the 12 code generation SLMs used for inference in this study. These models were selected through a systematic and reproducible process (see also~\ref{experimental_units}).}
        \label{tab:models_table}
    
    \resizebox{\textwidth}{!}{%
\begin{tabular}{cllrlrll} 
    \toprule 
    \makecell[c]{\textbf{Model Size} \\ \textbf{Category}} & \textbf{Model} & \textbf{Release Date} & \textbf{Size [M/B]} & \textbf{Affiliation} & \makecell[c]{\textbf{HumanEval} \\ \textbf{(Pass@1)}} & \makecell[c]{\textbf{Hugging Face} \\ \textbf{Name}} \\ 
    \midrule 
    \multirow{4}{*}{1} 
    & codeparrot-small & November, 2021 & 110 [M] & HuggingFace & 3.80\% & codeparrot/codeparrot-small \\ 
    & tiny-starcoder & December, 2023 & 164 [M] & BigCode & 7.84\% & bigcode/tiny\_starcoder\_py \\ 
    & pythia-410m & April, 2023 & 410 [M] & EleutherAI & 1.20\% & EleutherAI/pythia-410m \\ 
    & bloomz-560m & May, 2023 & 560 [M] & BigScience & 2.18\% & bigscience/bloom-560m \\ 
    \midrule
    \multirow{4}{*}{2} 
    & starcoderbase-1b & December, 2023 & 1 [B] & BigCode & 15.17\% & bigcode/starcoderbase-1b \\ 
    & bloomz-1b1 & May, 2023 & 1.1 [B] & BigScience & 2.63\% & bigscience/bloomz-1b1 \\ 
    & tinyllama & January, 2024 & 1.1 [B] & StatNLP Research Group & 9.15\% & TinyLlama/TinyLlama-1.1B-Chat-v1.0 \\ 
    & pythia1-4b & April, 2023 & 1.4 [B] & EleutherAI & 4.30\% & EleutherAI/pythia-1.4b \\ 
    \midrule
    \multirow{4}{*}{3} 
    & codegemma-2b & June, 2024 & 2 [B] & Google & 31.10\% & google/codegemma-2b \\ 
    & phi2 & December, 2023 & 2.7 [B] & Microsoft & 48.00\% & microsoft/phi-2 \\ 
    & stablecode-3b & August, 2023 & 3 [B] & Stability AI & 26.89\% & stabilityai/stablecode-instruct-alpha-3b \\ 
    & stablecode-3b-completion & August, 2023 & 3 [B] & Stability AI & 17.68\% & stabilityai/stablecode-completion-alpha-3b-4k \\ 
    \bottomrule 
\end{tabular}
}

\end{table*}

\subsection{Dataset construction} 
\label{dataset_construction}

Our input dataset was derived utilizing a methodology similar to those outlined by Fried \emph{et al.} and Bavarian \emph{et al.} \cite{fried2022incoder,bavarian2022efficient}. Specifically, we selected a subset of prompts from the HumanEval dataset \cite{chen2021evaluating} by taking each problem and processing it as follows: 
we tokenized the entire problem and then randomly selected a sequence of 10 to 15 tokens to form an input prompt. This token sequence was decoded and used as an input prompt for the model, with the aim of generating additional tokens based on the sampled input. 
By repeating this procedure across the entire HumanEval set, we created a collection of short‐prompt (fewer than 15 tokens) completion requests that are both controlled to fit our computational setup and diverse enough to effectively evaluate the SLMs in code generation tasks with respect to our dependent variables \cite{ilager2025green, sun2024neural}. The resulting dataset closely emulates interactive IDE‐style auto‐complete code task and ensures uniformity in prompt length when comparing inference efficiency across different serving configurations \cite{weber2024significant}.

Although our dataset is generated from the HumanEval benchmark, consistent with prior work that evaluates language models on this suite \cite{perrault2024artificial}, we concentrate exclusively on auto-complete code scenario rather than large-context or conversational code generation scenarios. Modern IDEs typically trigger completion suggestions after only a few tokens have been entered, making a 10 to 15 token window a realistic proxy for real-world usage \cite{jimenez2023swe}. Consequently, our dataset captures short-prompt use cases and does not cover tasks that involve navigating a large codebase, synthesizing entire functions across multiple files, or engaging in multi-turn conversational exchanges with an LLM, such as understanding how functions in different files connect or spotting a small bug in code evolving over time \cite{weber2024significant,jimenez2023swe}. While this choice restricts our findings to auto-complete task rather than full-function or deep conversational tasks, it provides a controlled environment for rigorously comparing serving behavior, such as execution time, energy consumption, and computing-resource utilization, when generating small code completions. All data and scripts for dataset creation are available in our replication package.

\subsection{Study design}

\begin{figure*}[htp]
    \centering\includegraphics[width=\textwidth]{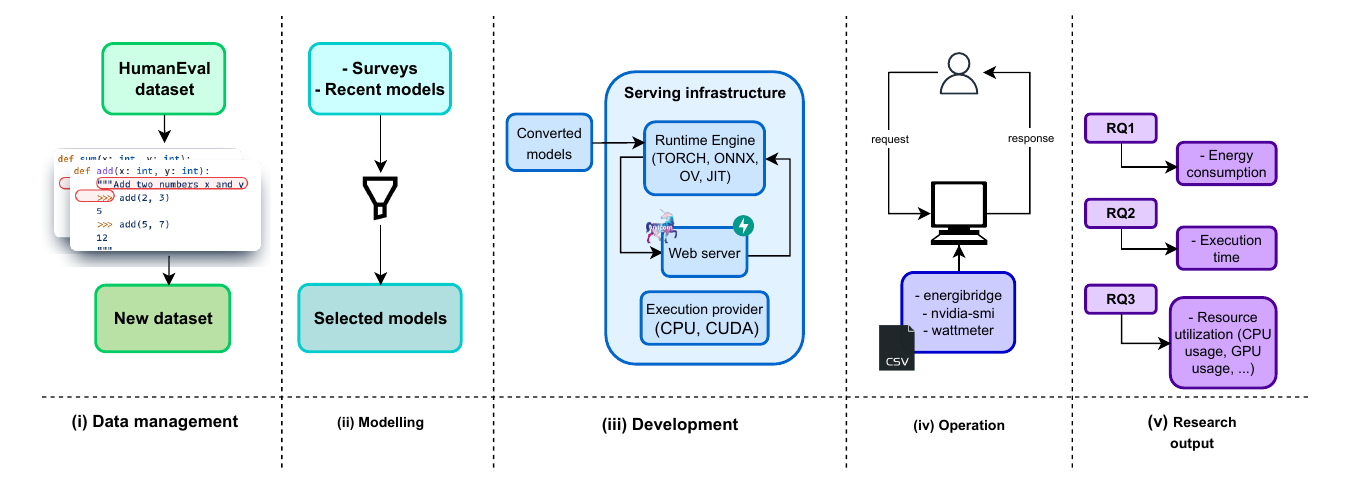}
    \caption{Schema of our experiment}
\label{fig:schema_study}
\end{figure*}

We conduct a technology-oriented experiment \cite{wohlin2012experimentation} following the guidelines by Jedlitschka \emph{et al.} \cite{jedlitschka2008reporting}. We selected twelve trained models and structured our investigation into a multi-stage pipeline, as illustrated in Figure~\ref{fig:schema_study}. This pipeline consists of five main stages \cite{lwakatare2020devops, castanyer2024design}:

\begin{enumerate}

    \item \textbf{Data management:}
    This foundational phase is critical in setting up our experiment to test code generation SLMs. We start with the HumanEval dataset \cite{chen2021evaluating}, which includes 164 coding problems, and adapt it to better suit the specific requirements of our study involving SLMs. This adaptation involves generating a new dataset through a token sampling strategy from HumanEval problems for the code generation task (see also~\ref{dataset_construction}). 
    Our study design leverages these generated prompts to evaluate multiple serving configurations, emulating a practical yet controlled environment. While HumanEval provides self-contained problems that reflect basic to intermediate coding challenges, it does not encompass, for instance, large-scale or multi-file code generation. We selected it for consistency with existing literature and ease of reproducibility \cite{perrault2024artificial}. 

    \item \textbf{Modelling:}
    Our study focuses on analyzing different configurations of deep learning serving infrastructures (see Subsection~\ref{independent_variables}), utilizing pre-trained models. This approach is in alignment with the objectives of our study (see also Subsection~\ref{experimental_units}).

    \item \textbf{Development:}
    This initial stage involves model conversion and the development of the serving infrastructure, which includes the REST API. Model conversion is aimed at adapting the pre-trained models to be compatible with the used runtime engines (see data availability statement in Section~\ref{sec:introduction}). 
    
    \item \textbf{Operation:}  
    After the development phase, the operation stage starts with the deployment of models across various serving configurations using a Uvicorn server\footnote{\url{https://www.uvicorn.org/}}, which is a high-speed ASGI web server and facilitates REST API requests via the FastAPI\footnote{\url{https://fastapi.tiangolo.com/}} API we developed. This setup allows real-time inferences with our input dataset.

    \item \textbf{Research Output:} The final phase analyzes the results generated in the previous stage, focusing on the dependent metrics. These outcomes allow us to answer our RQs posed in our study, according to each serving infrastructure configuration. Furthermore, these results may provide guidelines on the best selection of runtime engines and execution providers.

    
    
\end{enumerate}

\subsection{Variables} \label{variables}

In the following we define the variables of our experimental design grouped into two categories, as summarized in Table~\ref{tab:variables}.

\subsubsection{Independent variables} \label{independent_variables}

In this study we define one independent variable, Serving Configuration, which is a duplet of a runtime engine and an execution provider (see Section~\ref{sec:background}). The setup consists of evaluating different deep learning serving configurations. We initially obtain eight possible configurations by combining two execution providers (CPU and CUDA) with four runtime engines (TORCH, ONNX, OV, and JIT). However, we exclude the $\langle OV, CUDA \rangle$ configuration, as OV is specifically optimized for Intel hardware and does not have compatibility with the CUDA execution provider \cite{intelov_runtime}. As a result, we analyze seven configurations, each representing a unique combination of runtime engine and execution provider (e.g., $\langle ONNX, CUDA \rangle$).


\subsubsection{Dependent variables}

To measure the impact of our independent variable to resource utilization, the study tracks some of the most widely measured and relevant metrics in the field according to the Green Software Measurement Model \cite{guldner2024development}: the energy consumption, execution time, and computing-resource utilization (CPU usage, RAM usage, GPU usage, GPU memory utilization, and GPU used memory) during the inference (see also Table~\ref{tab:variables}).

\begin{table*}[t]
    \centering
        \caption{Variables of the experiment.}
        \label{tab:variables}
    
    \resizebox{\textwidth}{!}{
    \begin{tblr}{
      row{1} = {c},
      cell{2}{1} = {c},
      cell{3}{1} = {r=7}{c},
      hline{1-3,10} = {-}{},
      hline{4-9} = {2-5}{},
    }
    \textbf{Class} & \textbf{Name}          & \textbf{Description}                                                                                                         & \textbf{Scale} & \textbf{Operationalization}              \\
    Independent    & Serving configuration  & {Duplet of a runtime engine and an execution \\provider: $\langle TORCH, CPU \rangle$, $\langle ONNX, CPU \rangle$, \\$\langle OV, CPU \rangle$, \textcolor{gray}{$\langle JIT, CPU \rangle$}, $\langle TORCH, CUDA \rangle$, \\$\langle ONNX, CUDA \rangle$, and  \textcolor{gray}{$\langle JIT, CUDA \rangle$}.}                                                                     & Nominal        & See sections \ref{sec:background} and \ref{variables}.                          \\
    Dependent      & Energy consumption     & {Energy consumption of one serving configuration \\ during inference with input dataset (J).}                                     & Numerical      & {Profiled \\ (energibridge,\\nvidia-smi).} \\
                   & Execution time         & {Average of duration to do inference with input\\ dataset (s).}                                                                    & Numerical      & {Profiled\\ (linux).}                     \\
                   & CPU usage              & Average percentage of used CPU (\%).                                                                                          & Ratio          & {Profiled\\ (energibridge).}              \\
                   & RAM usage              & Average percentage of used RAM (\%).                                                                                              & Ratio          & {Profiled\\ (energibridge).}              \\
                   & GPU usage              & Average percentage of used GPU (\%).                                                                                              & Ratio          & {Profiled\\ (nvidia-smi).}                \\
                   & GPU memory utilization & {Average percentage of time over the past sample \\ period during which global (device) memory \\ was being read or written (\%).} & Ratio          & {Profiled\\ (nvidia-smi).}                \\
                   & GPU used memory        & {Average percentage of memory allocated by \\ active contexts (\%).}                                                               & Ratio          & {Profiled\\ (nvidia-smi).}                
    \end{tblr}
    }

\end{table*}

\subsection{Experiment setting} \label{experiment_setting}

All these experiments have been conducted with an NVIDIA GeForce RTX 4090 GPU with 24 GB memory with CUDA Version 12.4, and an AMD Ryzen 9 7950X 16-Core Processor CPU, 2.5 GHz frequency with 32 GB of
DDR5 5600 MHz RAM.

The complete experiment and data analysis is implemented in Python 3.10.12, using HuggingFace Transformers 4.40.2 \cite{wolf2020transformers}, and Optimum 1.19.2\footnote{\url{https://huggingface.co/docs/optimum/}}. To implement the serving configuration, we followed the implementations and model exportation provided by HuggingFace.

To accurately measure the variables outlined in our study, we use the energibridge profiler, nvidia-smi profiler, and a wattmeter. These tools are recognized in the scientific community for their effectiveness in assessing the resource utilization of computational processes, as evidenced by their application in previous research \cite{xu2023energy, martinez2024energy}. EnergiBridge was chosen for its simplicity and ready-to-use setup, enabling efficient data collection across various platforms without extensive configuration. Nvidia-smi, optimized for NVIDIA GPUs, was the ideal tool for measuring GPU-specific resource utilization. The wattmeter provided direct and precise measurements of total device energy consumption. It was used to verify that the energy reported by the CPU and GPU profilers matched the total energy consumption measured by the wattmeter, ensuring the accuracy and validity of our energy measurements. 
Finally, we used the Linux date command to record timestamps at the start and end of each experiment. These were cross-referenced with the timestamps recorded by the profiling tools to ensure alignment and to verify that the system was either idle or actively running inferences.


\textbf{EnergiBridge.}\footnote{\url{https://github.com/tdurieux/EnergiBridge}} EnergiBridge is a cross-platform tool designed to measure software resource utilization, focused on energy consumption, on various operating systems (Linux, Windows, and macOS) and CPU architectures (Intel, AMD, and Apple ARM). Developed in Rust, it simplifies resources monitoring by providing a unified interface and collects data through low-level system calls, converting measurements into watts for standardized reporting across different systems \cite{sallou2023energibridge}.

\textbf{NVIDIA System Management Interface (nvidia-smi).\footnote{\url{https://developer.nvidia.com/nvidia-system-management-interface}}} 
The nvidia-smi command-line tool is a practical utility for accessing a variety of NVIDIA GPU-related data, including power consumption, memory usage, temperature, and more. This tool is especially valuable for monitoring the power use of GPU-intensive tasks such as training deep learning models, inference tasks, streaming videos, gaming, and other similar activities.

\textbf{Wattmeter.}\footnote{\url{https://vitriko.eu/regleta-inteligente-netio-powerbox-4kf}} 
The wattmeter is a tool for measuring electrical power. In our experimental setup, we used the NETIO PowerBOX 4KF, an intelligent power strip with four individually controllable outlets. The experimental system was connected to the monitored outlets, enabling accurate measurement of key electrical metrics such as energy consumption. Measurements were retrieved through the device’s JSON-based API. The wattmeter was configured according to the manufacturer’s setup instructions to ensure consistency and reproducibility in all experiments.

\subsection{Experiment execution}
In this study, we deploy a server that receives HTTP requests from a client. The server hosts the developed API, which processes input prompts and returns the model's output, specifically the generated decoded tokens. The experiment involves selecting one of the seven serving configurations and performing the following steps for each configuration: for each model, the model is loaded, and inference is conducted on all 164 requests using the input dataset. During the inference process for each serving configuration, we collect the dependent variables using the two profilers described in Section~\ref{experiment_setting}. To ensure stability, we run the server for five minutes before starting the inferences and for another five minutes after completing the inferences with each serving configuration. The experiment is repeated 10 times to ensure consistency in the results, as this number of repetitions was sufficient to observe minimal variability and provide reliable medians across configurations. 
Still, each repetition of the experiment involves 7 serving configurations, 12 models, and 164 input requests per model, resulting in a substantial amount of data collected.


\subsection{Data analysis}\label{sec:data_analysis}

\begin{table*}[t]
    \centering
        \caption{Tests performed for RQs with their null hypotheses.}
        \label{tab:hypothesis}
    
    \begin{tblr}{
      cell{2}{3} = {r},
      cell{3}{3} = {r},
      cell{4}{1} = {r=5}{c},
      cell{4}{3} = {r},
      cell{5}{3} = {r},
      cell{6}{3} = {r},
      cell{7}{3} = {r},
      cell{8}{3} = {r},
      hline{1-4,9} = {-}{},
      hline{5-8} = {2-4}{},
    }
    \textbf{RQs} & \textbf{Variable}          & {\textbf{Statistical test/}\\\textbf{ Post-hoc test}} & \textbf{Null hypothesis}                                                                                                                  \\
    RQ1          & Energy consumption         & {Welch's ANOVA/\\Games-Howell}                        & {\textbf{H.1.0:} There is no difference in energy consumed during inference\\varying the runtime engine and execution provider.}          \\
    RQ2          & Execution time             & {Welch's ANOVA/\\Games-Howell}                        & {\textbf{H.2.0: }There is no difference in execution time during inference\\varying the runtime engine and execution provider.}           \\
    RQ3          & CPU usage                  & {Welch's ANOVA/\\Games-Howell}                        & {\textbf{H.3.1.0:} There is no difference in CPU usage during\\inference varying the runtime engine and execution provider.}              \\
                 & RAM usage                  & {Welch's ANOVA/\\Games-Howell}                        & {\textbf{H.3.2.0:} There is no difference in RAM usage during\\inference varying the runtime engine and execution provider.}              \\
                 & GPU usage                  & {Kruskal-Wallys/\\Dunn}                               & {\textbf{H.3.3.0:} There is no difference in GPU usage during\\inference varying the runtime engine and execution provider.}              \\
                 & {GPU memory\\ utilization} & {Kruskal-Wallys/\\Dunn}                        & {\textbf{H.3.4.0:} There is no difference in GPU memory utilization during\\inference varying the runtime engine and execution provider.} \\
                 & {GPU used\\ memory}        & {Kruskal-Wallys/\\Dunn}                               & {\textbf{H.3.5.0: }There is no difference in GPU used memory during\\inference varying the runtime engine and execution provider.}        
    \end{tblr}

\end{table*}

To address the research questions (RQs), we define the null hypotheses showed in Table~\ref{tab:hypothesis} for each dependent variable and its performed tests.

For each hypothesis, we follow a systematic analysis process: (1) use box plots to illustrate the distributions for each dependent variable, comparing between configurations using the same model; (2) assess if the measurements are normally distributed and have equal variances across the different treatments of each RQ. Utilize the Shapiro-Wilk test to check for the normality of the data.  To check the homogeneity of variances, we use a Levene test for equality of variances.
(3) assess the statistical significance (i.e., p-value) of the findings.

In our data analysis, we conducted two types of statistical tests based on the distribution of the dependent variables. For configurations where the data is normally distributed, we applied Welch's ANOVA to assess differences between serving configurations, since the dependent variables do not exhibit equality of variances, followed by the Games-Howell test for post-hoc analysis. For non-normally distributed data, we used the Kruskal-Wallis test, with Dunn's test as the post-hoc analysis to identify specific differences between groups. The details of the analysis and test results are available in the replication package for further examination (see data availability statement in Section~\ref{sec:introduction}).

\section{Results}\label{sec:results}

\begin{figure*}[htb]
    \centering 

    \begin{subfigure}[b]{0.47\textwidth} 
        \includegraphics[width=\textwidth]{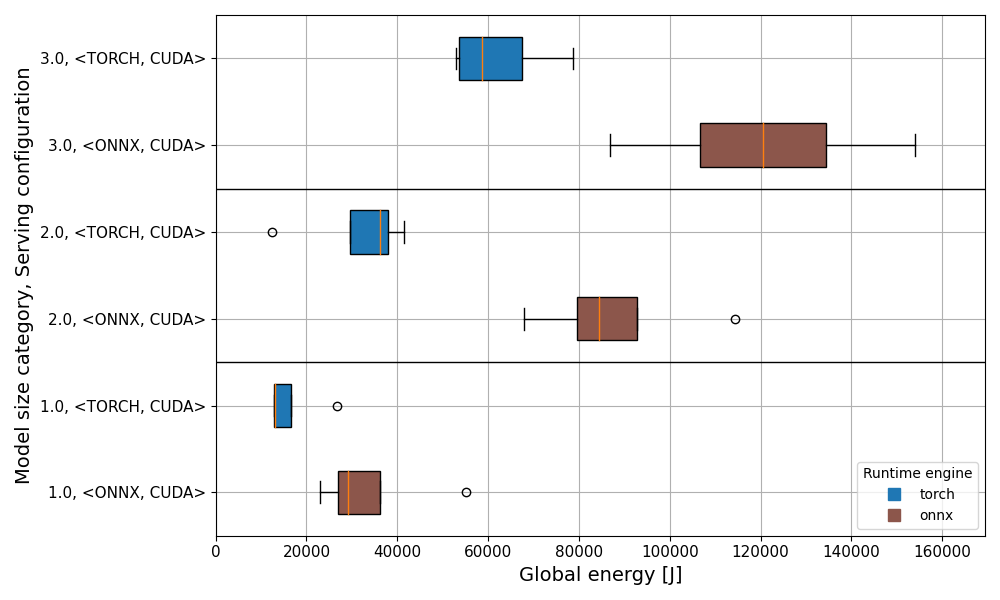}
        \caption{Energy consumption [J]}
        \label{fig:image1}
    \end{subfigure}
    \hfill 
    \begin{subfigure}[b]{0.47\textwidth}
        \includegraphics[width=\textwidth]{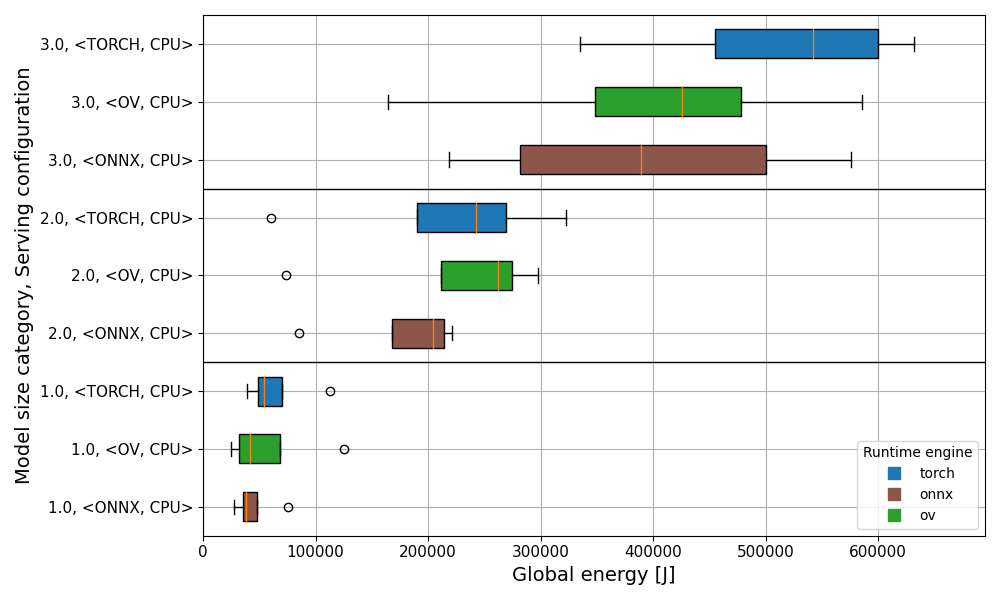}
        \caption{Energy consumption [J]}
        \label{fig:image2}
    \end{subfigure}

    \begin{subfigure}[b]{0.47\textwidth} 
        \includegraphics[width=\textwidth]{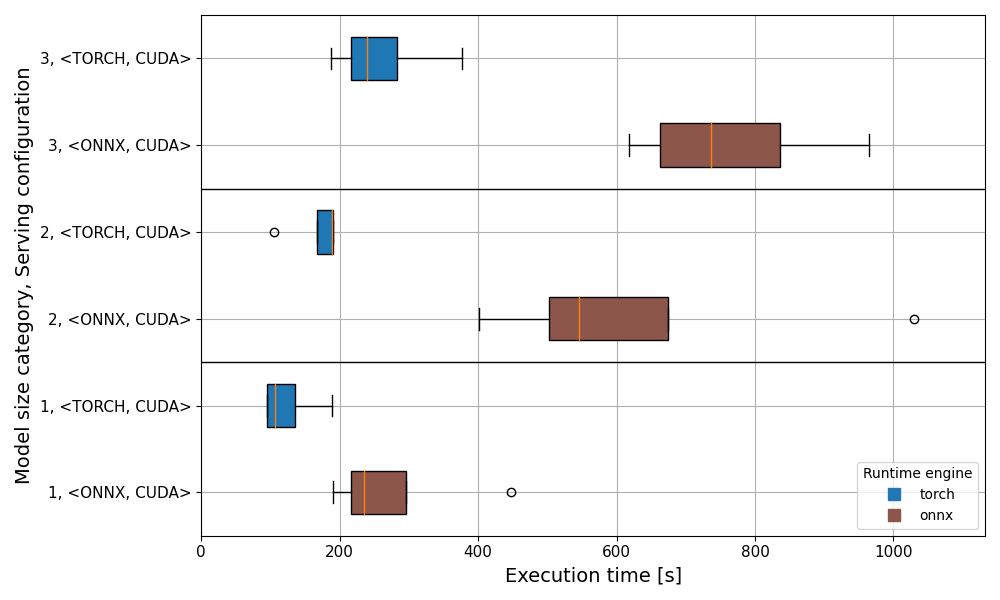}
        \caption{Execution time [s]}
        \label{fig:image1}
    \end{subfigure}
    \hfill 
    \begin{subfigure}[b]{0.47\textwidth}
        \includegraphics[width=\textwidth]{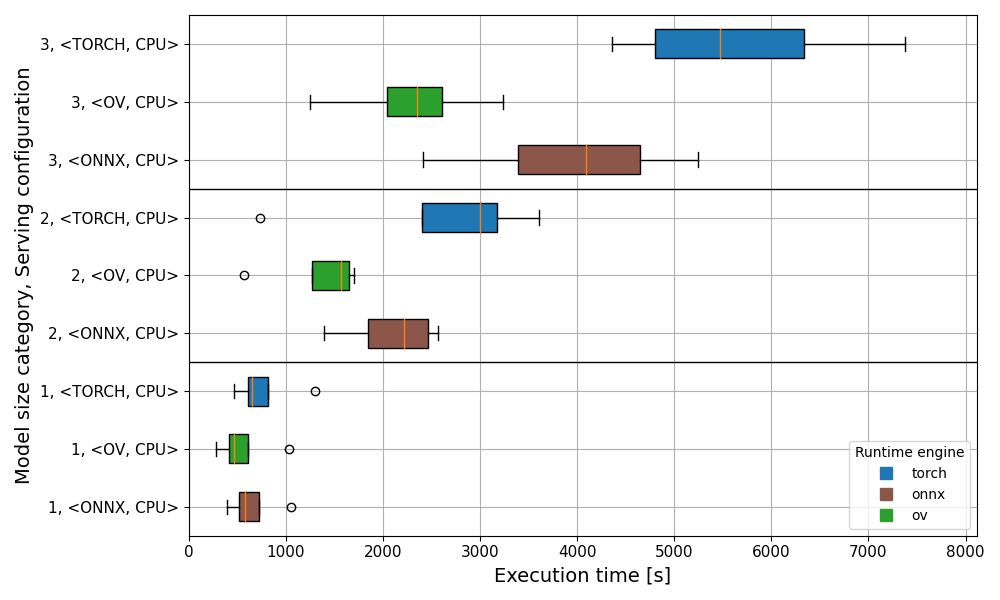}
        \caption{Execution time [s]}
        \label{fig:image2}
    \end{subfigure}
    
    \caption{Aggregated energy consumption and execution time by model size category: CUDA execution provider (left) and CPU execution provider (right) measured by Energibridge and Nvidia-smi}
    \label{fig:boxplots_energy_time}
\end{figure*}

\begin{figure*}[htb]
    \centering 

    \begin{subfigure}[b]{0.47\textwidth} 
        \includegraphics[width=\textwidth]{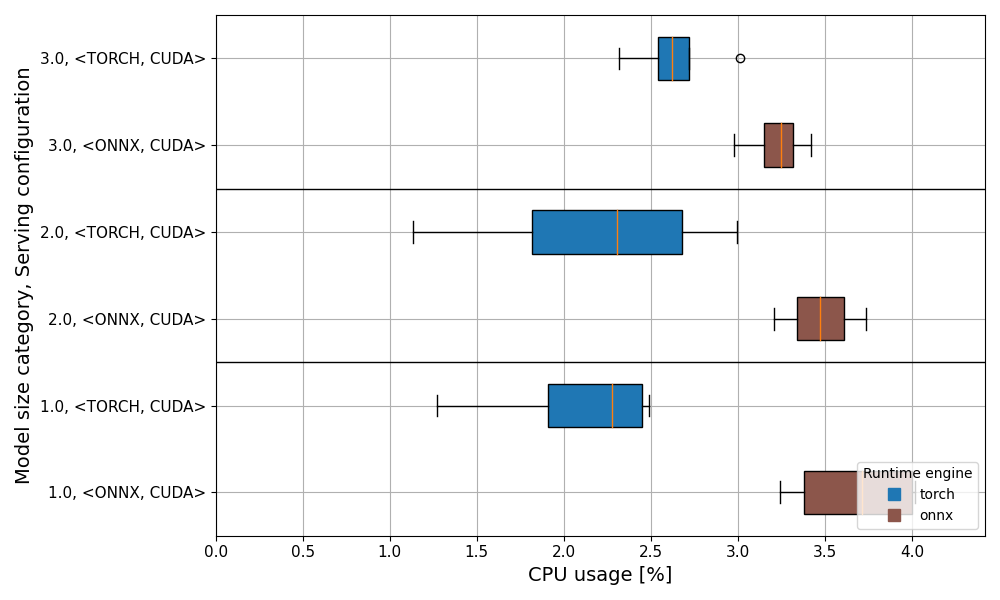}
        \caption{CPU usage [\%]}
        \label{fig:image1}
    \end{subfigure}
    \hfill 
    \begin{subfigure}[b]{0.47\textwidth}
        \includegraphics[width=\textwidth]{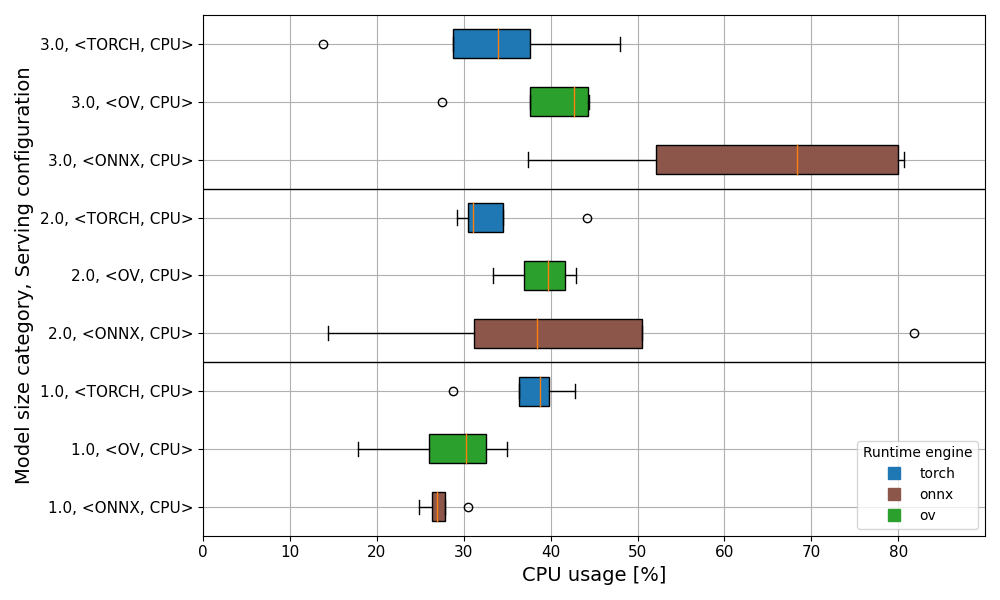}
        \caption{CPU usage [\%]}
        \label{fig:image2}
    \end{subfigure}

    \begin{subfigure}[b]{0.47\textwidth} 
        \includegraphics[width=\textwidth]{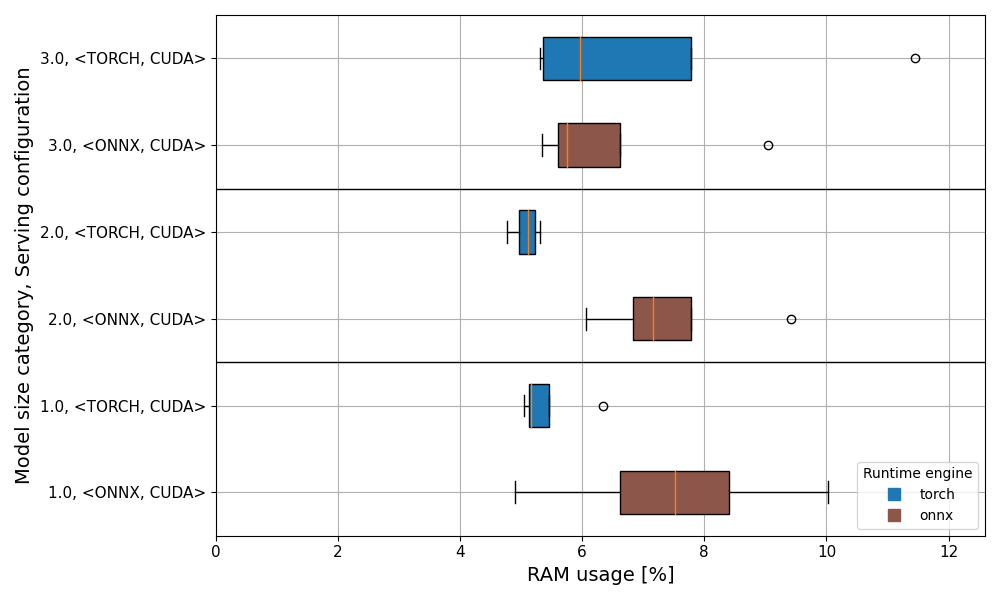}
        \caption{RAM usage [\%]}
        \label{fig:image1}
    \end{subfigure}
    \hfill 
    \begin{subfigure}[b]{0.47\textwidth}
        \includegraphics[width=\textwidth]{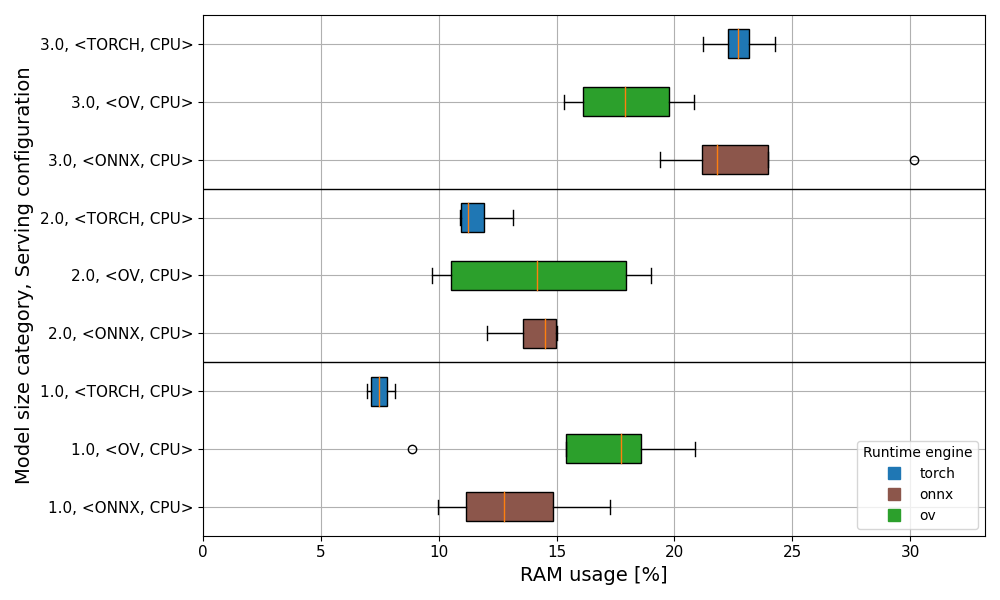}
        \caption{RAM usage [\%]}
        \label{fig:image2}
    \end{subfigure}
    
    \caption{Aggregated computing-resource utilization by model size category: CUDA execution provider (left) and CPU execution provider (right) measured by Energibridge}
    \label{fig:boxplots_cpu_ram}
\end{figure*}

\begin{figure*}[htb]
    \centering 

    \begin{subfigure}[b]{0.47\textwidth} 
        \includegraphics[width=\textwidth]{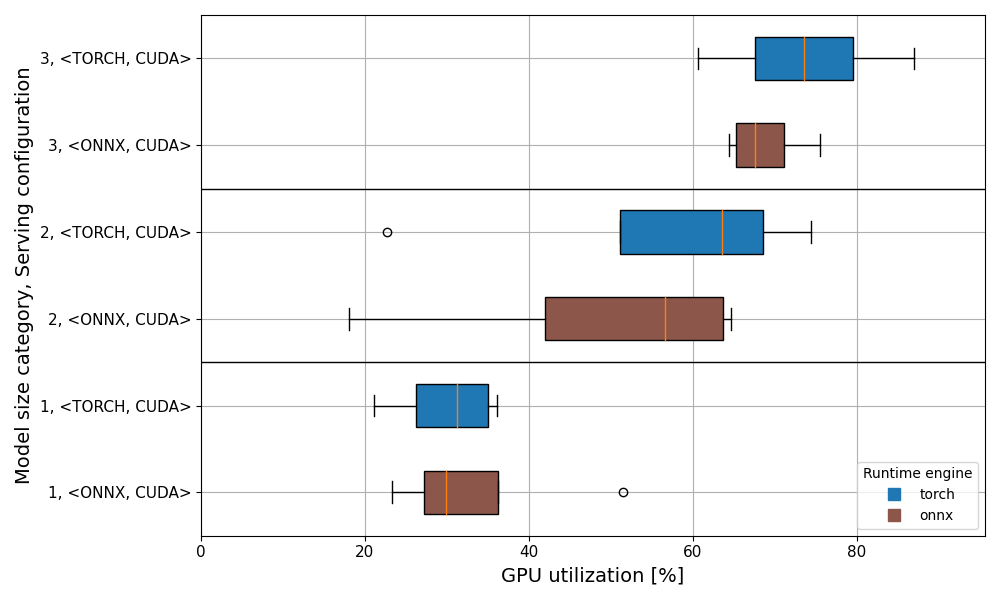}
        \caption{GPU utilization [\%]}
        \label{fig:image1}
    \end{subfigure}
    \begin{subfigure}[b]{0.47\textwidth}
        \includegraphics[width=\textwidth]{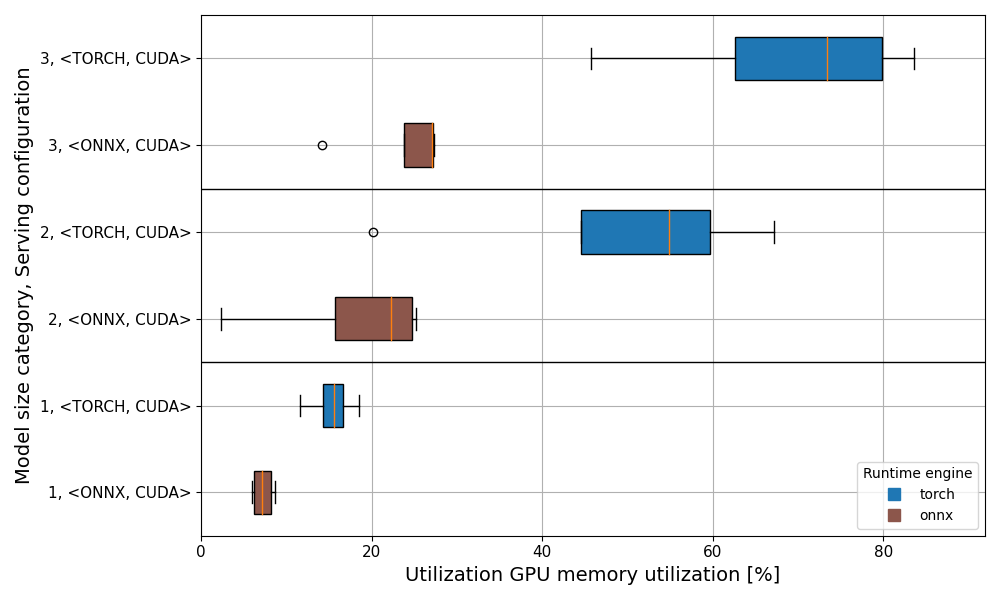}
        \caption{Utilization GPU memory utilization [\%]}
        \label{fig:image2}
    \end{subfigure}

    \vspace{0.5cm} 
    \begin{subfigure}[b]{0.47\textwidth}
        \includegraphics[width=\textwidth]{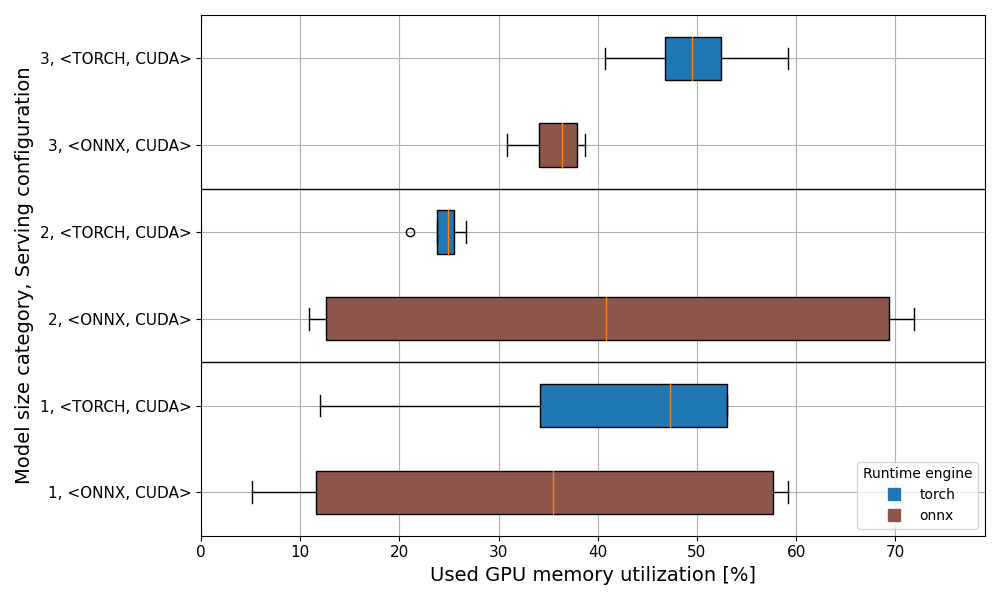}
        \caption{Used GPU memory utilization [\%]}
        \label{fig:image3}
    \end{subfigure}
    \hfill
    
    \caption{Aggregated computing-resource utilization by model size category: CUDA execution provider configurations measured by Nvidia-smi}
    \label{fig:boxplots_nvidia}
    
\end{figure*}

In this section, we report the results of our experiment, addressing our research questions and highlighting key takeaways. We initially evaluated seven deep learning serving configurations. Following the experiments, we reviewed the outputs of the scripts and inferences, identifying error-prone responses in the JIT configurations, $\langle JIT, CPU\rangle$ and $\langle JIT, CUDA \rangle$. 
JIT’s Trace mode generated error-prone responses because it records a fixed computational graph from sample inputs and cannot account for conditions outside the traced path. Given these issues, and to preserve fairness in our comparisons, we excluded JIT configurations ($\langle JIT, CPU\rangle$ and $\langle JIT, CUDA \rangle$) from further analysis.
The results presented here focus on comparisons across the final five selected configurations while using results from twelve different code generation SLMs, as shown in Tables~\ref{tab:energy_time_results_aggregated_cpuep} -~\ref{tab:energy_time_results_aggregated_cudaep} and Figures~\ref{fig:boxplots_energy_time} -~\ref{fig:boxplots_nvidia} (one color for each model in the box plots).

\begin{figure*}[htb]
    \centering 

    \begin{subfigure}[b]{0.47\textwidth} 
        \includegraphics[width=\textwidth]{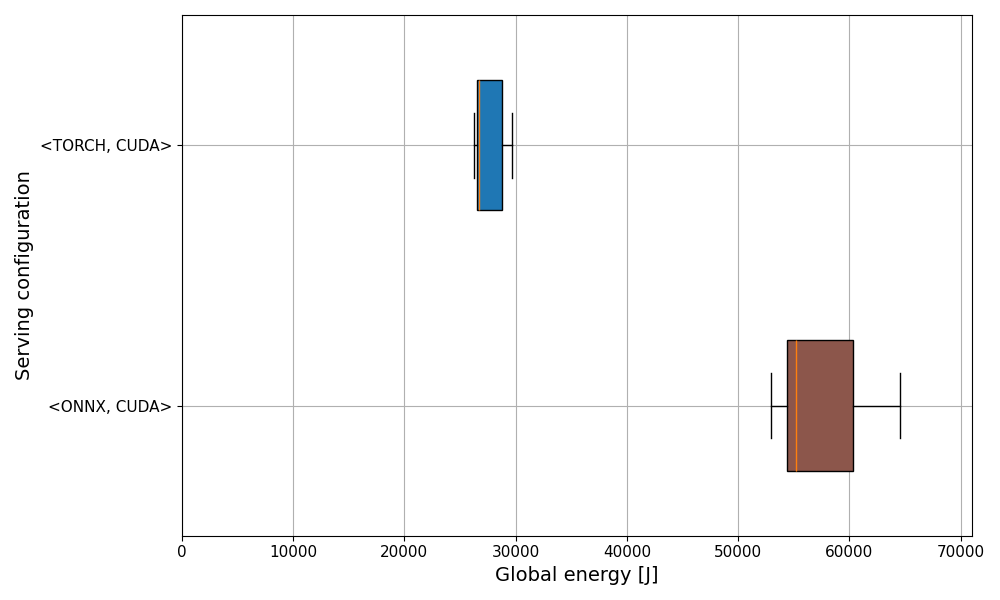}
        \caption{Energy consumption [J]}
        \label{fig:image1}
    \end{subfigure}
    \hfill 
    \begin{subfigure}[b]{0.47\textwidth}
        \includegraphics[width=\textwidth]{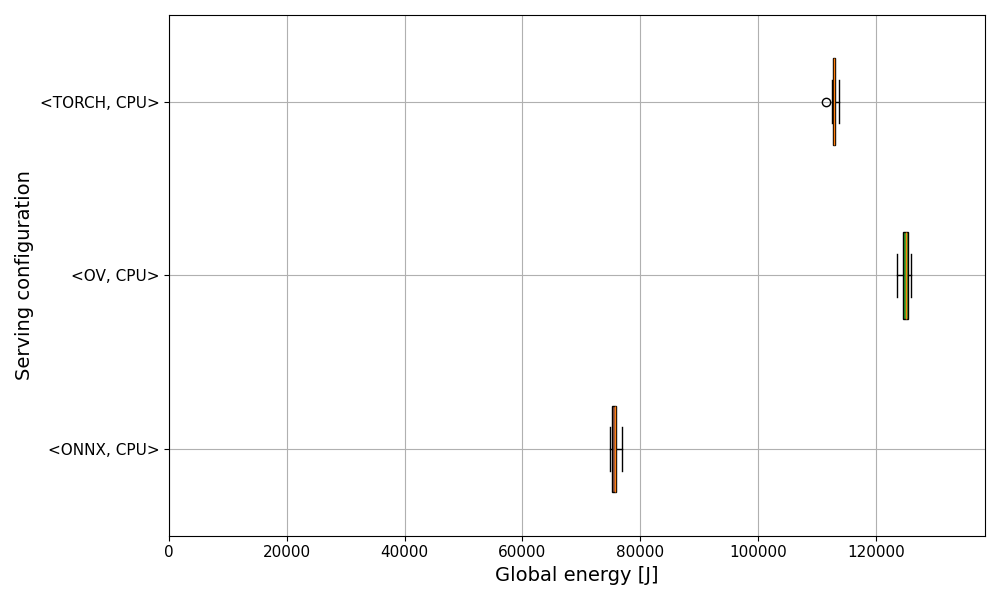}
        \caption{Energy consumption [J]}
        \label{fig:image2}
    \end{subfigure}

    \begin{subfigure}[b]{0.47\textwidth} 
        \includegraphics[width=\textwidth]{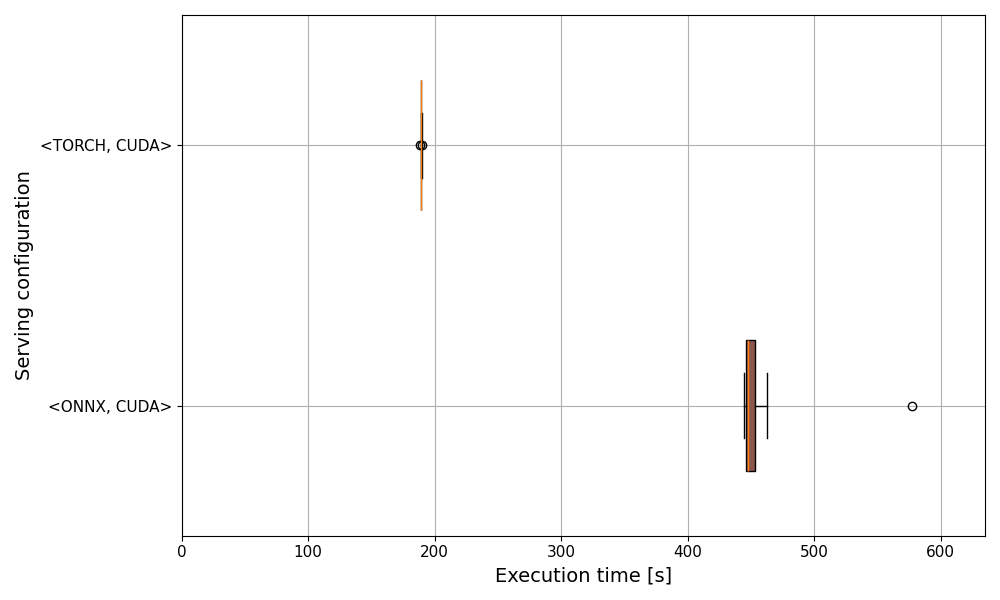}
        \caption{Execution time [s]}
        \label{fig:image1}
    \end{subfigure}
    \hfill 
    \begin{subfigure}[b]{0.47\textwidth}
        \includegraphics[width=\textwidth]{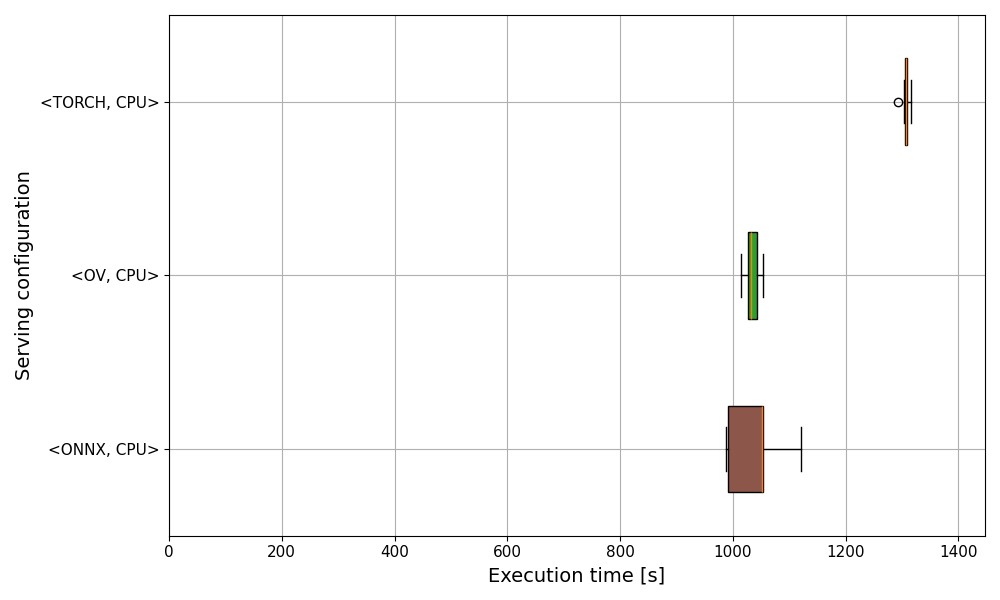}
        \caption{Execution time [s]}
        \label{fig:image2}
    \end{subfigure}
    
    \caption{Example of energy consumption and execution time for model \textit{pythia-410m} before model size category aggregation: CUDA execution provider (left) and CPU execution provider (right) measured by Energibridge and Nvidia-smi}
    \label{fig:dis_boxplots_energy_time}
\end{figure*}

\begin{figure*}[htb]
    \centering 

    \begin{subfigure}[b]{0.47\textwidth} 
        \includegraphics[width=\textwidth]{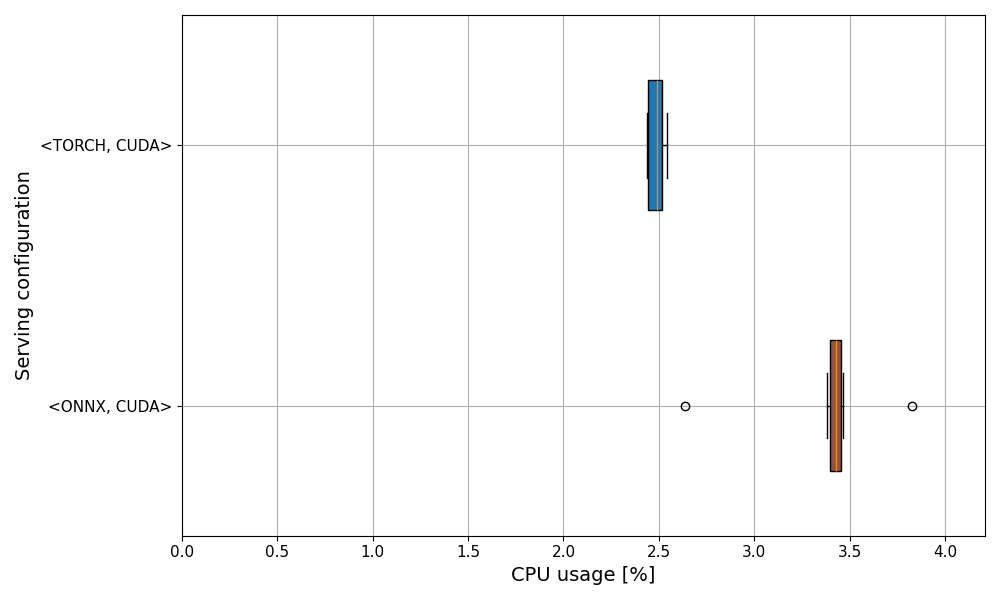}
        \caption{CPU usage [\%]}
        \label{fig:image1}
    \end{subfigure}
    \hfill 
    \begin{subfigure}[b]{0.47\textwidth}
        \includegraphics[width=\textwidth]{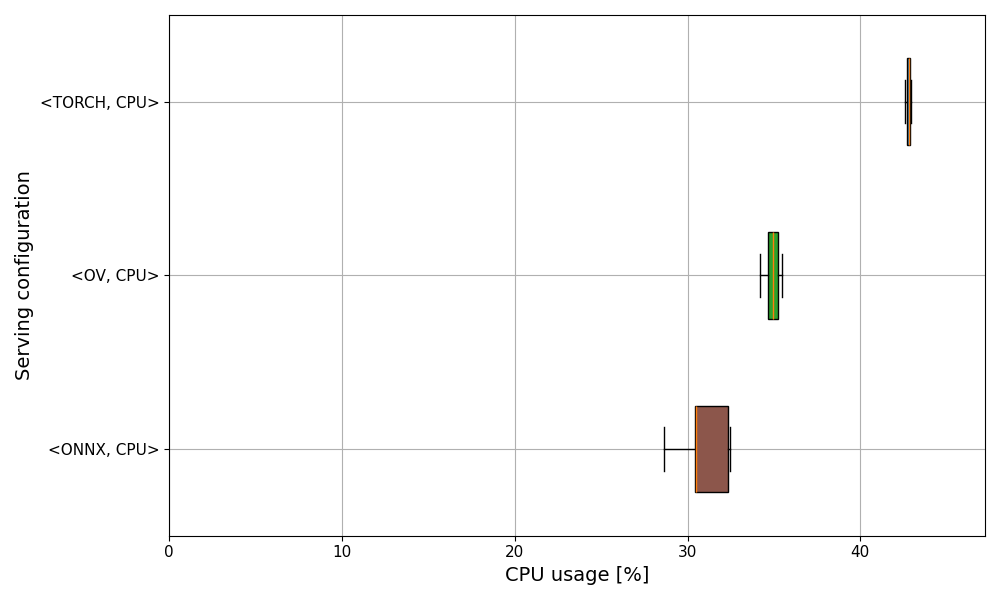}
        \caption{CPU usage [\%]}
        \label{fig:image2}
    \end{subfigure}

    \begin{subfigure}[b]{0.47\textwidth} 
        \includegraphics[width=\textwidth]{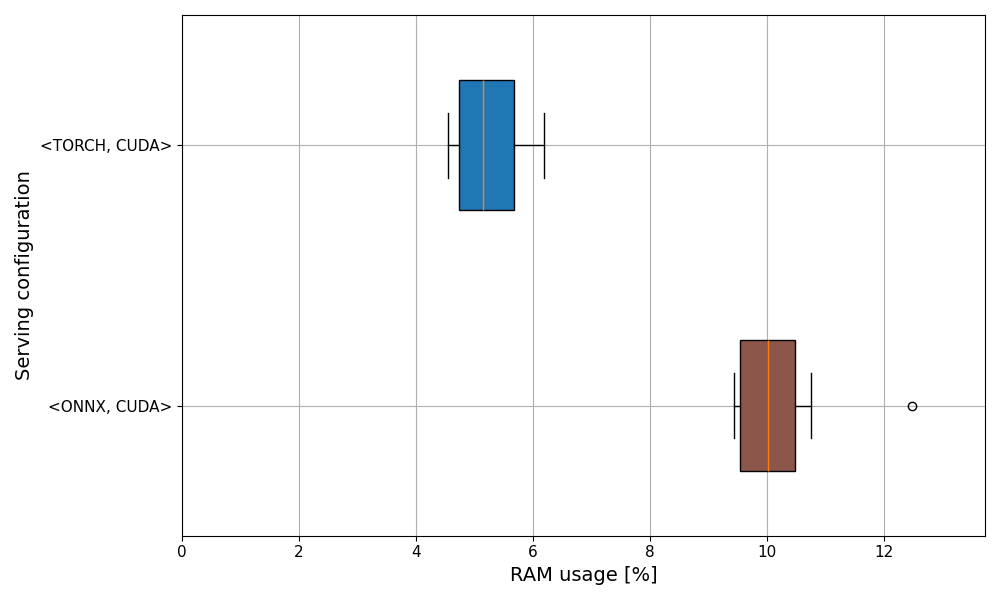}
        \caption{RAM usage [\%]}
        \label{fig:image1}
    \end{subfigure}
    \hfill 
    \begin{subfigure}[b]{0.47\textwidth}
        \includegraphics[width=\textwidth]{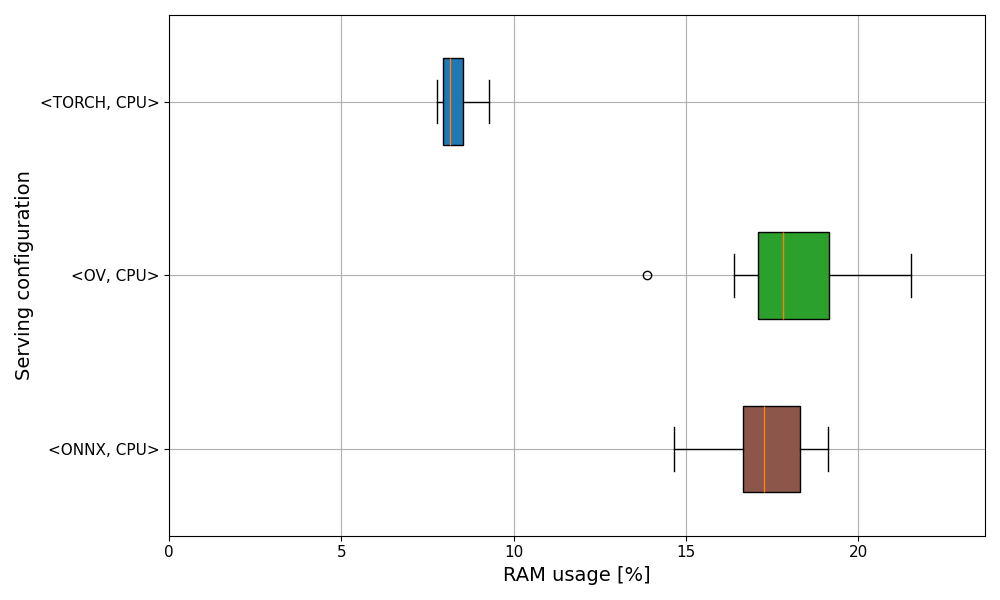}
        \caption{RAM usage [\%]}
        \label{fig:image2}
    \end{subfigure}
    
    \caption{Example of computing-resource utilization for model \textit{pythia-410m} before model size category aggregation: CUDA execution provider (left) and CPU execution provider (right) measured by Energibridge}
    \label{fig:dis_boxplots_cpu_ram}
\end{figure*}

\begin{figure*}[htb]
    \centering 

    \begin{subfigure}[b]{0.47\textwidth} 
        \includegraphics[width=\textwidth]{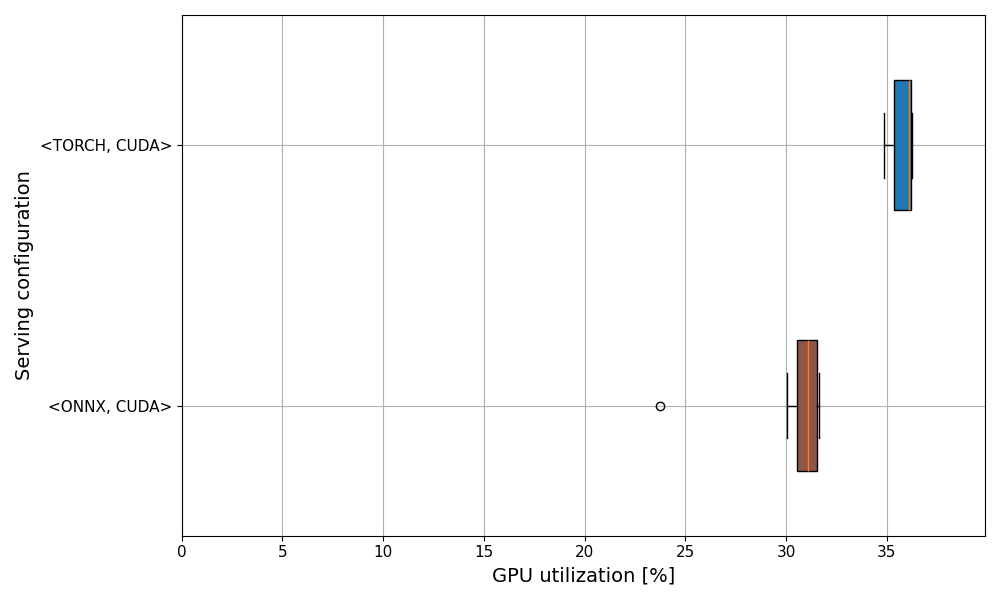}
        \caption{GPU utilization [\%]}
        \label{fig:image1}
    \end{subfigure}
    \begin{subfigure}[b]{0.47\textwidth}
        \includegraphics[width=\textwidth]{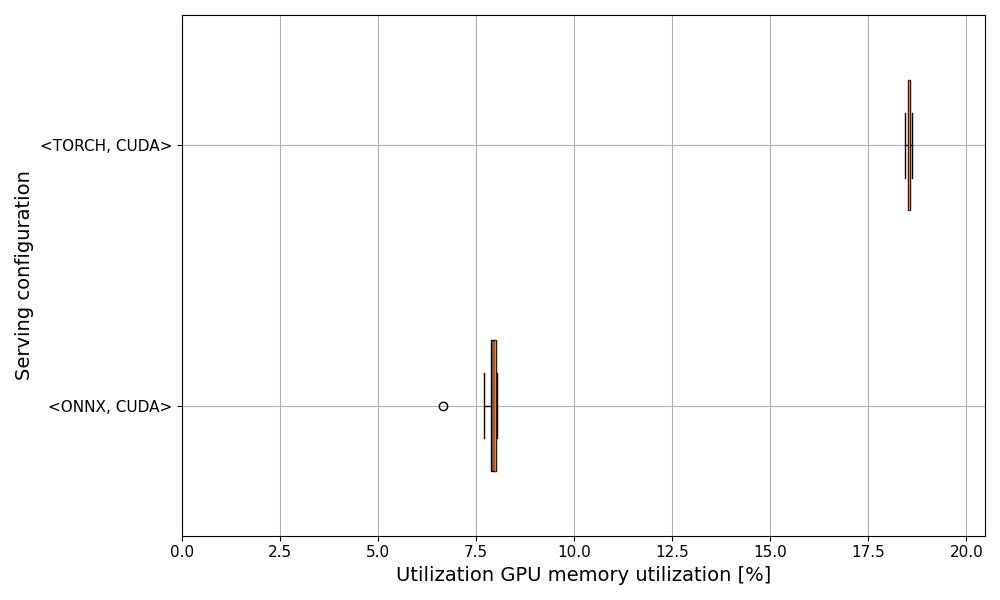}
        \caption{Utilization GPU memory utilization [\%]}
        \label{fig:image2}
    \end{subfigure}

    \vspace{0.5cm} 
    \begin{subfigure}[b]{0.47\textwidth}
        \includegraphics[width=\textwidth]{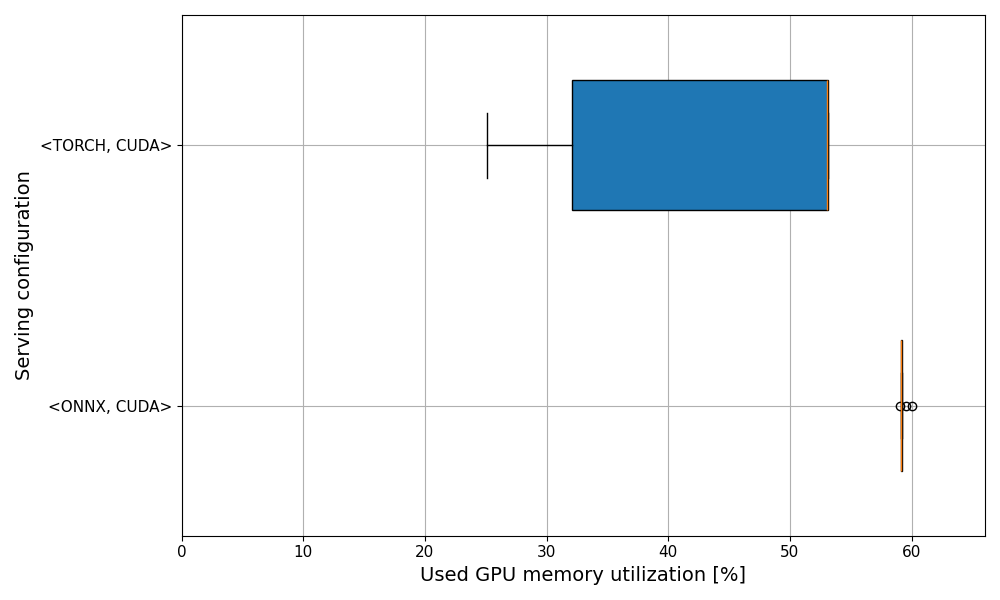}
        \caption{Used GPU memory utilization [\%]}
        \label{fig:image3}
    \end{subfigure}
    \hfill
    
    \caption{Example of computing-resource utilization for model \textit{pythia-410m} before model size category aggregation: CUDA execution provider configurations measured by Nvidia-smi}
    \label{fig:dis_boxplots_nvidia}
    
\end{figure*}

In some figures, as in Figure~\ref{fig:boxplots_energy_time}, we grouped four different models in size category 3, each with its own mean values. That grouping inevitably produces greater overall dispersion than looking at any one of those models separately. Consequently, the difference stems from aggregating models with distinct baselines and performance profiles into a single category rather than reflecting instability in our measurements.
As an example, Figures~\ref{fig:dis_boxplots_energy_time}–\ref{fig:dis_boxplots_nvidia} present box plots for the model \textit{pythia-410m}, shown before aggregation by model size category. These additional figures illustrate execution time, energy consumption, and computing-resource utilization across all serving configurations for a single model, showing the low variability observed during the experiment.
Furthermore, the replication package contains disaggregated plots for all individual models, which consistently exhibit similarly low variability. 



Due to the extensive amount of experimental data generated from assessing twelve distinct models across five serving configurations, we aggregated results into clearly defined model-size categories (see Table~\ref{tab:models_table}). The categories are organized based on model size category, with category 1 representing the smallest models and category 3 representing the largest models. Median values were specifically chosen to summarize categories due to their robustness against outliers, given substantial variability observed within size categories. 
Meanwhile, mean values were employed to summarize serving configurations owing to their more uniform distributions, facilitating direct comparative analysis across configurations.

It is important to explicitly note that the detailed model-by-model analysis was primarily conducted at the model level (without aggregation), and remains fully available in our replication package. Aggregated data here serves solely as a concise, high-level representation of overall trends.

\subsection{Energy consumption of code generation SLMs(RQ1).}


Tables~\ref{tab:energy_time_results_aggregated_cpuep} and~\ref{tab:energy_time_results_aggregated_cudaep} present the aggregated median values of energy consumption by model size category. For CPU execution provider configurations (where the GPU is not utilized), only global energy, based on CPU energy, is reported. For CUDA execution provider configurations, the results include CPU energy, GPU energy, and global energy, calculated as the sum of CPU and GPU energy.


The median energy consumption for configurations using the CUDA execution provider is consistently lower than that of the CPU execution provider. Notably, the $\langle TORCH, CUDA \rangle$ configuration stands out as the most energy-efficient across all models, a trend that is also evident in the aggregated results.

The $\langle ONNX, CUDA \rangle$ configuration exhibits the highest median energy consumption values among the CUDA execution provider configurations. Similarly, $\langle TORCH, CPU\rangle$ configurations consume the most energy across the majority of models when using the CPU execution provider. This pattern is also reflected in the aggregated results.

ONNX optimizations are significantly more effective when paired with the CPU execution provider, demonstrating better energy efficiency compared to the deep learning framework alone. Among the optimized runtime engines, ONNX along with OV, ONNX emerges as the most energy-efficient in CPU execution provider configurations. However, when using the CUDA execution provider, ONNX fails to provide the same level of energy savings, as it does not outperform the baseline TORCH runtime engine, which remains the most energy-efficient. This discrepancy in energy performance likely stems from the way the ONNX runtime engine interacts with each execution provider.

Although execution time alone is not a sufficient proxy for energy consumption, as discussed later, the substantial differences observed between CPU and CUDA configurations suggest that extended inference durations likely contribute to increased energy use in CPU-based settings (see Table~\ref{tab:energy_time_results_aggregated_cpuep}).
This prolonged inference time, especially with more complex models, may cause CPU configurations to use substantially more energy. The inefficiency of the CPU execution provider in energy conservation highlights the need for optimized resource allocation and management in deep learning serving environments to minimize unnecessary energy consumption. In contrast, CUDA configurations, despite utilizing both CPU and GPU resources, were far more energy-efficient for intensive tasks due to their significantly faster execution times.


Our results showcase considerable differences in the energy consumption of different configurations. There is statistic significance (Welch's ANOVA test p-value) between the deep learning serving configuration and the response variables. Furthermore, Games-Howell post-hoc test reveals that all configurations are statistically different from each other (\textbf{H.1.0}). These results are provided in the replication package (see data availability statement in Section~\ref{sec:introduction}).

\begin{tcolorbox}
    

    \textbf{Answer to RQ1:} Energy consumption is significantly affected by the serving configuration, with CUDA configurations being more efficient than CPU configurations. $\langle TORCH, CUDA \rangle$ provides the best energy efficiency, while optimized runtime engines like OV and ONNX, particularly ONNX, reduce energy consumption in CPU configurations but are less effective with CUDA compared to TORCH.

\end{tcolorbox}

\begin{table*}[t]
    \centering
        \caption{CPU execution provider: aggregated median energy consumption and execution time values for inference across entire dataset by model size category. Best values are shown in green. Last section presents mean values for each serving configuration.}
        \label{tab:energy_time_results_aggregated_cpuep}

    \begin{threeparttable}
\begin{tabular}{rlrr}
\toprule
Model size category & Serving configuration & Global energy [J] & Execution time [s] \\
\midrule
1 & $\langle ONNX, CPU \rangle$ & \cellcolor{green!30}{45069.35} & 655.09 \\
1 & $\langle OV, CPU \rangle$ & 58374.88 & \cellcolor{green!30}{561.61} \\
1 & $\langle TORCH, CPU \rangle$ & 64993.75 & 773.27 \\
\midrule
2 & $\langle ONNX, CPU \rangle$ & \cellcolor{green!30}{178641.05} & 2100.83 \\
2 & $\langle OV, CPU \rangle$ & 223821.62 & \cellcolor{green!30}{1355.18} \\
2 & $\langle TORCH, CPU \rangle$ & 216854.71 & 2583.76 \\
\midrule
3 & $\langle ONNX, CPU \rangle$ & \cellcolor{green!30}{393187.23} & 3958.18 \\
3 & $\langle OV, CPU \rangle$ & 400527.50 & \cellcolor{green!30}{2300.14} \\
3 & $\langle TORCH, CPU \rangle$ & 512731.24 & 5668.91 \\
\specialrule{0.2em}{0.5em}{0.5em} 
1,2,3 & $\langle ONNX, CPU \rangle$ & \cellcolor{green!40}{205632.54} & 2238.03 \\
1,2,3 & $\langle OV, CPU \rangle$ & 227574.67 & \cellcolor{green!40}{1405.64} \\
1,2,3 & $\langle TORCH, CPU \rangle$ & 264859.90 & 3008.65 \\
\bottomrule
\end{tabular}

    \end{threeparttable}
\end{table*}

\begin{table*}[t]
    \centering
        \caption{CUDA execution provider: aggregated median energy consumption and execution time values for inference across entire dataset by model size category. Best values are shown in green. Last section presents mean values for each serving configuration.}
        \label{tab:energy_time_results_aggregated_cudaep}

    \begin{threeparttable}
    \resizebox{\textwidth}{!}{
\begin{tabular}{rlrrrr}
\toprule
Model size category & Serving configuration & CPU energy [J] & GPU energy [J] & Global energy [J] & Execution time [s] \\
\midrule
1 & $\langle ONNX, CUDA \rangle$ & 13404.99 & 20692.81 & 34138.28 & 277.12 \\
1 & $\langle TORCH, CUDA \rangle$ & 5193.32 & 11250.87 & \cellcolor{green!30}16432.53 & \cellcolor{green!30}124.03 \\
\midrule
2 & $\langle ONNX, CUDA \rangle$ & 32113.95 & 55652.21 & 87788.33 & 630.64 \\
2 & $\langle TORCH, CUDA \rangle$ & 7287.23 & 24247.47 & \cellcolor{green!30}31525.72 & \cellcolor{green!30}168.74 \\
\midrule
3 & $\langle ONNX, CUDA \rangle$ & 39012.21 & 81516.82 & 120509.17 & 763.59 \\
3 & $\langle TORCH, CUDA \rangle$ & 12196.07 & 50052.11 & \cellcolor{green!30}62301.62 & \cellcolor{green!30}260.62 \\
\specialrule{0.2em}{0.5em}{0.5em} 
1,2,3 & $\langle ONNX, CUDA \rangle$ & 28177.05 & 52620.61 & 80811.93 & 557.12 \\
1,2,3 & $\langle TORCH, CUDA \rangle$ & 8225.54 & 28516.82 & \cellcolor{green!40}36753.29 & \cellcolor{green!40}184.46 \\
\bottomrule
\end{tabular}
}
    \end{threeparttable}
\end{table*}

\subsection{Execution time of code generation SLMs (RQ2).}

Tables~\ref{tab:energy_time_results_aggregated_cpuep} and~\ref{tab:energy_time_results_aggregated_cudaep} present the aggregated median values of execution time by model size category.  This analysis highlights a significant difference in execution time, with configurations utilizing the CUDA execution provider generally achieving faster execution times compared to those using CPU execution provider. This underscores the efficiency of GPU acceleration in handling computationally intensive tasks.


Extreme cases are observed with models such as phi2 using $\langle TORCH, CPU \rangle$, where the median execution time (5983.52 seconds) is more than 20 times longer than with the $\langle TORCH, CUDA \rangle$ configuration (251.97 seconds), highlighting significant bottlenecks in CPU-based processing for complex computational tasks. Similar extreme cases are noted within the largest model size category (Category 3) using TORCH, where the CPU execution provider’s execution time (5668.91 seconds) exceeds the CUDA execution provider’s time (260.62 seconds) by more than 20 times.




Similar to its impact on energy efficiency, ONNX optimizations are clearly more effective with the CPU execution provider, but not as successful with the CUDA execution provider, comparing to Torch configurations. When using the CPU execution provider, the optimized runtime engines, particularly OV and ONNX, perform faster than the deep learning framework alone, with OV showing the best results. However, ONNX behaves differently when paired with the CUDA execution provider, where it fails to match or outperform TORCH in terms of speed. Again, this discrepancy may be due to how each execution provider interacts with the ONNX runtime engine. This results can also be observed in Tables~\ref{tab:energy_time_results_aggregated_cpuep} and~\ref{tab:energy_time_results_aggregated_cudaep}.

Our experiments revealed that while execution time optimization often aligns with reduced energy consumption, there exist notable exceptions. Specifically, within CPU configurations, the $\langle OV, CPU \rangle$ configuration exhibited faster execution times compared to $\langle ONNX, CPU\rangle$ but consistently required higher energy consumption. This explicitly shows scenarios where optimizing time performance alone may not directly lead to energy efficiency.

Welch’s ANOVA confirmed significant differences in execution times across different runtime engines. The subsequent Games-Howell post-hoc tests annotate that all configurations are statistically distinct from one another, confirming the marked impact of the execution provider and runtime engine choice on the efficiency of computational tasks (\textbf{H.2.0}). The only exceptions were that the execution times of the $\langle ONNX, CPU\rangle$ and $\langle TORCH, CPU\rangle$ configurations did not differ significantly for three out of the twelve models (bloomz-560m,  codeparrot-small, and tiny-starcoder), resulting in the rejection of \textbf{H.2.0} for only these models. 

\begin{tcolorbox}
    

    \textbf{Answer to RQ2:} Execution time is significantly influenced by the serving configuration, with CUDA configurations consistently outperforming CPU configurations. $\langle TORCH, CUDA \rangle$ delivers the best performance, while OV and ONNX optimize execution time, especially OV, with CPU execution provider but do not achieve comparable improvements with CUDA execution provider, reflecting similar patterns seen in energy consumption.
    
\end{tcolorbox}

        
        

\subsection{Computing-resource utilization of code generation SLMs (RQ3).}

Table~\ref{tab:hypothesis} presents a summary of the statistical analysis conducted to assess the computing-resource utilization across the different serving configurations. These analyses were applied to key metrics such as CPU usage, RAM usage, GPU usage, GPU memory utilization, and GPU used memory. 


For statistical evaluation, different tests were applied depending on the normality and variance homogeneity of the data. Welch’s ANOVA was used for data with unequal variances but normal distributions, while the Kruskal-Wallis test was employed for non-normally distributed data. The significance of differences between configurations was assessed using these tests, with p-values indicating significant differences with values below $\alpha = 0.05$. Furthermore, post-hoc analyses using Games-Howell or Dunn’s test were performed to identify specific pairwise differences. In terms of computing-resource utilization metrics, all serving configurations showed significant differences from one another (\textbf{H.3.1.0, H.3.2.0, H.3.3.0, H.3.4.0, and H.3.5.0}), with a few exceptions. Specifically, RAM usage between certain serving configuration pairs—such as $\langle OV, CPU\rangle$ and $\langle TORCH, CPU\rangle$, $\langle ONNX, CUDA \rangle$ and $\langle TORCH, CUDA \rangle$, $\langle OV, CPU\rangle$ and $\langle ONNX, CPU\rangle$, as well as $\langle TORCH, CPU\rangle$ and $\langle ONNX, CUDA \rangle$—did not exhibit significant differences in two out of the twelve models for each pair. As a result, \textbf{H.3.2.0} was rejected for those specific cases.


In the results of GPU metrics, we focus exclusively on the three CUDA execution provider configurations. Comparisons with CPU execution provider configurations are omitted, as they do not utilize the GPU, making such comparisons irrelevant.

We profiled computing-resource utilization  using described profiling tools, capturing explained metrics. VRAM (GPU memory) was only utilized in configurations involving the CUDA execution provider. In CPU-only configurations, VRAM usage remained consistently negligible or zero, as expected. These results confirm the distinction in memory handling strategies across execution providers.

\textbf{CPU usage:} The visual data and median values indicate that configurations using the CPU execution provider generally exhibit higher CPU usage than those employing the CUDA execution provider, which aligns with expectations given the operational differences between the two. For CUDA execution providers, CPU bottlenecks are minimal since the majority of computations are offloaded to the GPU. Conversely, CPU-based configurations concentrate the workload entirely on the processor and system memory, resulting in higher RAM and CPU usage.


The $\langle TORCH, CUDA \rangle$ configuration consistently showed the lowest CPU usage among all configurations, indicating that it effectively shifts computation to the GPU wherever possible. By reducing reliance on the CPU and leveraging GPU capabilities for parallel operations, this setup optimizes computing-resource utilization.


CPU usage for configurations utilizing the ONNX runtime engine is slightly higher when paired with the CUDA execution provider, with median values ranging from 2.98\% to 4.02\%. While differences in CPU usage are observed, no consistent pattern emerges across models when the CPU execution provider is employed.



\textbf{RAM usage:} Results show that configurations using the CPU execution provider generally consume more RAM compared to those using the CUDA execution provider. This discrepancy can be attributed to the fact that CPU execution provider manages all computational resources directly within the system's main memory, whereas CUDA execution provider leverages both RAM and dedicated VRAM, effectively distributing memory load.

The $\langle TORCH, CUDA \rangle$ configuration consistently demonstrated the lowest RAM usage, ranging from 4.77\% to 11.45\% in 9 out of 12 models. These percentages correspond to the cases where $\langle TORCH, CUDA \rangle$ ranked lowest for RAM usage among all GPU-based configurations. In the remaining 3 models, a different GPU-based configuration achieved the lowest RAM usage. This highlights its efficiency in computing-resource utilization, as it achieves minimal energy and execution time, and minimal RAM consumption. TORCH configurations, in general, exhibit lower RAM usage compared to other runtime engines.

In contrast, $\langle ONNX, CPU\rangle$ and $\langle OV, CPU\rangle$ configurations exhibited significantly higher RAM usage, with $\langle ONNX, CPU\rangle$ ranging from 9.96\% to 30.16\% and $\langle OV, CPU\rangle$ from 8.85\% to 20.88\% in 10 out of 12 models. These findings indicate that optimized runtime engines like ONNX and OV demand more system resources due to their memory management strategies.
While they improve energy efficiency and execution time in some scenarios, these benefits often involve higher RAM usage. In scenarios with limited computational resources, increased memory requirements can indeed become a significant concern, outweighing gains in execution time or energy savings. Furthermore, in CUDA execution provider configurations, ONNX exhibits higher RAM usage but performs worse in terms of both time and energy efficiency compared to TORCH, further underscoring TORCH's superior results of performance in all three dependent metrics.


\textbf{GPU usage:} results show that the configuration $\langle TORCH, CUDA \rangle$ reaches the highest levels of GPU utilization (from 21.08\% up to 86.98\%), indicating that a substantial portion of the available GPU compute capacity is actively used during inference. Conversely, the $\langle ONNX, CUDA \rangle$ configuration shows the lowest GPU usage (from 18.07\% up to 75.49\% ) among the CUDA execution provider options, suggesting it may not exploit the computational potential of the GPU as efficiently as $\langle TORCH, CUDA \rangle$. This trend is also reflected in the aggregated results across the three categories.



\textbf{GPU memory utilization:} Our GPU utilization memory results reveal that $\langle TORCH, CUDA \rangle$ configurations exhibited the highest memory utilization (from 11.57\% up to 83.59\%, all models), effectively leveraging GPU capabilities for intensive memory operations. In contrast, $\langle ONNX, CUDA \rangle$ configurations showed the lowest memory utilization (from 2.34\% up to 27.33\%), suggesting conservative number of GPU VRAM operations compared to configurations using $\langle TORCH, CUDA \rangle$.


These findings emphasize the crucial role of runtime engine selection in maximizing the memory efficiency of CUDA execution provider, with significant variations noted in how different engines manage and utilize VRAM.



\textbf{GPU used memory:} Regarding GPU allocated memory (GPU used memory), the results show varying patterns among CUDA execution provider configurations. For 7 out of 12 models, the $\langle TORCH, CUDA \rangle$ configuration exhibited higher levels of memory allocation, while the $\langle ONNX, CUDA \rangle$ configuration showed higher memory allocation for the remaining 5 models. Specifically, $\langle TORCH, CUDA \rangle$ configurations ranged from 12.01\% to 59.16\%, whereas $\langle ONNX, CUDA \rangle$ configurations ranged from 5.12\% to 71.87\%. These findings highlight differences in memory allocation strategies.

\begin{tcolorbox}

    \textbf{Answer to RQ3:} 

    Computing-resource utilization is significantly influenced by the serving configuration. CPU-based configurations (particularly with OV or ONNX) exhibit higher RAM usage, reflecting increased memory demands. In contrast, CUDA configurations, particularly $\langle TORCH, CUDA \rangle$, show more efficient computing-resource utilization, achieving higher GPU utilization and better memory allocation, reducing the load on CPU and RAM.
\end{tcolorbox}

        
        


\section{Discussion and implications}\label{sec:discussions}

In this section, we analyze and interpret the study's findings for each dependent variable, discuss their implications, and highlight key insights derived from the research.

%
\textbf{Energy consumption.}
In our analysis, we observe that configurations utilizing the CUDA execution provider consistently consume less energy compared to those utilizing the CPU execution provider. This significant difference underscores the inherent efficiencies of CUDA execution provider, which enables hardware-accelerated computation on NVIDIA GPUs, effectively managing parallel processing tasks in code generation SLMs.


Interestingly, ONNX runtime configurations show greater optimization when paired with CPU execution provider compared to CUDA execution provider, when compared to other runtime engines under the same execution provider. This enhanced performance with CPU execution provider could stem from the specific libraries used (such as Transformers and Optimum HuggingFace's libraries \cite{wolf2020transformers}) or the particular demands of code generation SLMs. It can be resource-intensive for ONNX configurations because of the overhead involved in mapping different subgraphs to suitable execution providers and the synchronization of inputs \cite{bekhelifi2024optimizing}. However, this complex binding process could yield benefits in scenarios involving varied batch sizes or other types of deep learning tasks, where such dynamic resource allocation is advantageous. 


Although $\langle TORCH, CUDA \rangle$ registers the lowest total energy in our experiments, two runtime engine artifacts make its lead over $\langle ONNX, CUDA \rangle$ look larger than it might truly be. First, limited kernel coverage: several decoder-specific operators lack CUDA implementations in the evaluated ONNX Runtime release, so they fall back to the CPU execution provider, raising host load and data transfers \cite{nvidia2024gtc, stiles2024efficient}. Second, sub-optimal I/O binding (IOBinding): every operator executed on the CPU incurs host-device and device-host copies, further incrementing energy use \cite{onnxruntime2024iobinding}. The runtime engine logs (\textit{``24 Memcpy nodes added'', ``Some nodes not assigned to the preferred EP''}) confirm both the extra transfers and CPU fall-back for some nodes. Practitioners should therefore enable IOBinding, review verbose EP-placement logs and verify that critical nodes execute on the intended hardware before drawing conclusions.
Further exploration of this type of runtime engine across various deep learning tasks could provide valuable insights for software engineers.

The findings from our study highlight crucial implications for deploying deep learning models, particularly in how different configurations affect resource utilization. The superior energy efficiency of CUDA execution provider configurations suggests significant cost savings and reduced environmental impact, making GPU-accelerated environments preferable for intensive parallel processing tasks. Furthermore, the trade-offs observed with ONNX's varied optimization across execution providers, underline the importance of aligning runtime engines with the appropriate hardware to reduce energy consumption, without requiring changes to the model architecture or applying model-level optimizations such as model compression \cite{shi2022compressing}.

\textbf{Execution time.}
CUDA execution provider configurations consistently outpace those using CPU execution provider, demonstrating the substantial advantages of leveraging NVIDIA GPUs for parallel processing tasks. This performance disparity is consistent with findings from other studies, further confirming the superior processing speeds of GPUs \cite{van2024survey}.

The capabilities of CUDA execution provider configurations demonstrate their efficiency in both time and energy during inference processes, independent of the chosen runtime engine. This efficiency is consistently evident across all CUDA execution provider configurations, making it a highly effective choice for operating SLM for code generation, enhancing both speed and energy efficiency.

The ONNX runtime engine shows a significant difference in optimization between CPU execution provider and CUDA execution provider configurations. It performs well under CPU execution provider, likely due to tailored optimizations for CPU usage, but does not translate these benefits as effectively to CUDA execution provider, when comparing to the other runtime engines. We find this counterintuitive, given the common expectation that GPU implementations typically outperform CPU ones. We believe this may be due to two main factors: (i) the CUDA execution provider in ONNX runtime does not fully support CUDA kernels used for the SLMs \cite{singh2024impact}, and (ii) our selected models or batch size may not have been complex enough to fully exploit GPU parallelism, or we may not have identified the most optimal CUDA-specific configuration within the used library \cite{sever2021performance}. These findings underscore the importance of aligning software optimizations with the characteristics of the target hardware.
Due to the speed of development of deep learning research and development, we expect that this type of optimized runtime engines are, in general, a better option for real-world deep learning systems that require efficient resource utilization.

The median time advantage of $\langle TORCH, CUDA \rangle$  over $\langle ONNX, CUDA \rangle$ is also related to kernel available operators for CUDA execution provider and data transfers. Profiling showed that the remaining difference was driven by host–device \textit{memcpy} nodes inserted by the ONNX optimizer and operations falling back on the CPU execution provider. In other words, the gap can be largely an artifact of kernel-support and data transfer overheads rather than an intrinsic limitation of ONNX Runtime.
Our results favor $\langle TORCH, CUDA \rangle$ partly because experiments ran on a single NVIDIA GeForce RTX 4090, a GPU for which TORCH has targeted optimization due to its common use as deep learning framework. ONNX Runtime, by contrast, still lacked CUDA kernels for some decoder-centric kernels in the used library version. On other execution providers or future ONNX releases with full kernel parity and IOBinding well enabled, the ranking could invert \cite{stiles2024efficient}. We therefore recommend reviewing logs and verify optimizations like IOBinding are being used to exploit the faster kernels for the used hardware.

In their study, Bekhelifi \emph{et al.} \cite{bekhelifi2024optimizing} found that configurations using $\langle ONNX, CPU\rangle$ outperformed those using $\langle OV, CPU\rangle$ in terms of execution time for Brain-Computer Interface applications. For example, they observed that $\langle ONNX, CPU\rangle$ resulted in fewer backend calls compared to $\langle OV, CPU\rangle$ when profiling inference execution, suggesting that ONNX may execute a more streamlined inference. However, our results contrast with this finding, as $\langle OV, CPU\rangle$ proved to be faster in our experiments. This discrepancy highlights how the performance of optimized runtime engines can vary significantly across different deep learning tasks. Similarly, Pochelu \emph{et al.} reported that OV consistently outperformed other CPU-based serving configurations for CNN workloads, while emphasizing that this choice should align with both application requirements and hardware characteristics \cite{pochelu2024mastering}. 

We recommend further investigation into how these engines perform across a broader range of deep learning systems to better understand their task-specific efficiencies and limitations. Additionally, we recommend that software engineers assess inference performance within their specific deployment contexts and take advantage of the rapid experimentation capabilities offered by the tools in our replication package to identify the most suitable configuration for their needs.

Optimizing execution time frequently aligns with reduced energy consumption due to shortened computational workloads. However, this alignment is explicitly not universal. Certain configurations, notably $\langle OV, CPU\rangle$, optimized execution time yet showed increased energy demands compared to slower configurations like $\langle ONNX, CPU\rangle$, clearly demonstrating that faster execution can occasionally entail greater energy costs. Consequently, we explicitly advocate against using execution time as a sole proxy for energy efficiency. Instead, direct monitoring of energy consumption is recommended to avoid misinterpreting or overlooking discrepancies among the key metrics for the specific deep learning system.


\textbf{Computing-resource utilization.}
Configurations using CPU execution provider typically encounter bottlenecks in CPU and RAM usage. This underscores the need for careful selection and optimization of runtime engines to avoid overloading the CPU, as computational tasks in this execution provider heavily rely on RAM. In contrast, CUDA execution provider configurations benefit from the additional use of VRAM, distributing the memory load more effectively between system RAM and GPU memory.
For instance, ONNX and OV configurations, while energy-efficient in CPU-based configurations, require substantial memory (RAM and/or GPU memory) to handle inference. Using lower precision can help manage GPU memory when using CUDA execution provider \cite{singh2024impact}.

Within the CUDA execution provider configurations, significant statistical differences were observed in GPU utilization and memory efficiency. The $\langle TORCH, CUDA \rangle$ configuration consistently demonstrated the highest GPU usage, GPU memory utilization, and GPU used memory, indicating a greater reliance on GPU resources to perform inference computations. In this context, higher GPU utilization reflects the configuration’s ability to effectively offload workload from the CPU to the GPU, which is particularly desirable for computationally intensive tasks. In contrast, the $\langle ONNX, CUDA \rangle$ configuration showed comparatively lower GPU utilization, suggesting potential inefficiencies in how it distributes computation or interacts with specific code generation SLMs. Despite its higher GPU resource usage, $\langle TORCH, CUDA \rangle$ proved more efficient overall, as it consumed less CPU and RAM and achieved superior energy and execution time metrics. This suggests a balanced and resource-conscious allocation strategy that fully leverages available GPU capabilities.

\textbf{JIT runtime.}
As outlined in Section~\ref{sec:results}, the JIT runtime engine was also included in our experiments. However, we excluded its results from the analysis due to the error-prone outputs it produced. Instead, we investigated the causes of these issues, focusing on underlying factors rather than setup problems.

As a runtime engine, JIT's primary mode of operation in this study was through Trace mode, which inherently supports a broader range of Python features by operating beneath the Python layer. This mode records all PyTorch operations executed with a representative input to create an optimized TorchScript Intermediate Representation (IR). While this mode is advantageous for accommodating all Python features and ensuring consistent execution paths, it is notably error-prone as it does not capture dynamic control flows inherent in Python, it only supports functions with a single path, leading to inaccuracies in conditions where execution paths vary \cite{lopes2023torchy, ansel2024pytorch, torchscriptJIT}.

JIT's ability to fuse operations into larger computational blocks offers advantages by reducing memory reads and writes, thereby optimizing certain computational tasks. Additionally, intermediate results can be cached directly in registers or shared memory, minimizing the need for repeated memory access \cite{pytorchDiscussion, chen2024evt}. However, despite these memory optimizations, overall memory usage remains high, indicating that the benefits do not necessarily translate into lower memory consumption. This limitation, combined with JIT’s inability to effectively manage dynamic execution paths, diminishes its potential energy and speed advantages. These findings highlight the critical need to align runtime engine capabilities with the specific demands and infrastructure of the deployment environment to achieve optimal performance and computing-resource utilization.

If a static graph exported shows good responses, it could be the best option for a production environment, however it must be really well tested and it is time-exhaustive. There are some options being built that can eliminate this drawbacks \cite{lopes2023torchy, ansel2024pytorch}.

Despite its exclusion from our final set of serving configurations, JIT remains an interesting runtime engine thanks to its operation-fusion capabilities and reduced overhead in certain scenarios. Specifically, when a model’s inputs remain stable or do not require complex branching and thus do not rely on newly generated outputs being fed back into the input (as often happens in code generation) JIT can deliver notable performance gains \cite{goldfarb2024evaluating}.

Our follow-up investigation revealed that the Trace mode we used struggles with the dynamic control flows central to code generation tasks. If an input prompt triggers a path beyond the initially traced computational graph, the resulting outputs may be incomplete or incorrect. This limitation reflects JIT’s reliance on a fixed execution graph, which does not adapt well to unexpected input variations. These observations align with prior work indicating JIT’s lower robustness in settings where the execution path frequently changes \cite{lopes2023torchy, ansel2024pytorch, torchscriptJIT}.

While JIT remains an optimal solution in scenarios featuring minimal branching and stable inputs, it requires further refinement to reliably support more dynamic workloads, such as code generation.

\textbf{Guidelines and implications for selecting appropriate serving infrastructure.}
Determining the best configuration largely depends on the specific requirements and constraints of the deep learning serving system. Based on the results presented in Section~\ref{sec:results}, several key guidelines can be drawn for software engineers. These guidelines provide insights into selecting appropriate execution providers and runtime engines based on specific operational needs:

\begin{itemize}
\item \textbf{Battery or environmentally constrained serving} (e.g., edge inference): We recommend prioritizing energy consumption. CUDA execution provider configurations, particularly $\langle TORCH, CUDA \rangle$, are the most energy-efficient for code generation tasks. When GPU access is limited, CPU execution provider paired with ONNX runtime offers a balanced and low-energy alternative, making them ideal for energy-aware, non-GPU environments.

\item \textbf{Latency constrained serving} (e.g., real-time code completion in IDEs): We suggest prioritizing execution time. CUDA execution provider configurations consistently deliver faster performance across the board. Among these, $\langle TORCH, CUDA \rangle$ offers the fastest performance. In CPU configurations, OV provides the best execution time enhancements.

\item \textbf{Computing-resource constrained serving} (e.g., legacy systems): 
In CPU execution provider configurations, if RAM is a constraint, we suggest using $\langle TORCH, CPU \rangle$ as a more computing-resource balanced option, although it may not be optimal in terms of energy efficiency or execution time. For better overall resource management, $\langle TORCH, CUDA \rangle$ stands out among CUDA execution provider configurations, as it effectively leverages GPU resources while reducing the workload on the CPU and minimizing RAM usage.

\end{itemize}

This study contributes practical insights that can support software engineers in the design, deployment, and maintenance of resource efficient deep learning systems. By incorporating non-functional properties, such as energy consumption, execution time, and computing-resource utilization, into the evaluation of serving infrastructure, our methodology enables practitioners to make informed architectural decisions during system design and operation.

Specifically, engineers can replicate or adapt our experimental framework to analyze the resource utilization of their own deep learning inference under different infrastructure configurations. These insights can guide choices in runtime engines and execution providers, depending on the constraints and priorities of a given application context (e.g., energy efficiency, latency, or hardware limitations).

Moreover, our findings highlight the architectural impact of serving infrastructure components, which are often overlooked in early design phases. As shown in this study and in prior work \cite{duran2024identifying}, such components can significantly influence resource utilization related properties, underscoring the need to include them in architectural planning.

Finally, system-level monitoring of non-functional metrics can serve not only as a benchmarking tool but also as a continuous feedback mechanism in production environments. This supports ongoing system improvements and helps teams maintain alignment with operational and environmental goals over time.

\textbf{Optimizing language models inference.} 
While scaling up the size of language models has consistently been shown to improve their accuracy \cite{kaplan2020scaling}, it also introduces significant computational, memory, and environmental costs that can hinder practical deployment. In contrast, SLMs offer a more efficient alternative, particularly suited for code generation tasks, as they provide acceptable performance with lower resource demands. 
In this context, our study, along with prior research \cite{duran2024identifying}, emphasizes that optimizing serving configurations plays a central role in improving inference resource utilization. These system-level decisions can enhance resource utilization without modifying the model or relying on techniques such as model compression, offering a practical approach for the sustainable deployment of SLMs.

The increasing adoption of language models in real-world applications highlights the critical role of system-level decisions. In addition to runtime engines and execution providers, serving infrastructures such as DL-specific software (e.g., TorchServe) and end-to-end cloud services (e.g., AWS SageMaker) represent system-level optimizations that can enhance energy and performance efficiency without compromising model accuracy \cite{duran2024identifying, zhou2024survey}. These decisions align with the software engineering perspective, where employing green software tactics can effectively optimize deep learning systems throughout their lifecycle \cite{martinez2024environmental}.

Additionally, data-level optimizations (e.g., Retrieval-Augmented Generation) and model-level optimizations (e.g., model compression and dynamic inference) might complement system-level strategies in enhancing efficiency. For instance, dynamic neural networks, which adapt computational resources based on input complexity, exemplify a balanced approach to achieving both efficiency and effectiveness. Existing implementations of these techniques provide valuable insights while highlighting opportunities for further analysis, development, and broader implementations \cite{han2021dynamic, zhou2024survey, wang2024model}.

\section{Limitations and threats to validity}\label{sec:threats}

Our study has various limitations, primarily due to its focus on open-source SLMs, specific runtime engines, and execution providers. The findings may not generalize to other model types or serving configurations. Although the provided replication package allows for other models and configurations to be easily added, the conclusions are limited to the examined setup and code generation SLMs context. Our findings may not generalize to all deep learning systems, but they offer a foundation for optimizing and analyzing resource utilization-efficient serving configurations.
These results are specific to the particular used hardware, described in Section~\ref{experiment_setting}, and future work could mitigate this by incorporating a wider range of hardware, including different GPUs and CPUs, to enhance the robustness of the findings. 

Additionally, this study focused on the code generation task using prompts derived from the HumanEval dataset \cite{chen2021evaluating}. While HumanEval is one of the most widely used evaluation datasets, its scope may not fully capture the diversity of real-world scenarios. For instance, prompts in these type of scenarios can span multiple files, integrate external libraries, or involve different releases. This limitation is particularly relevant given that our research work and development of serving infrastructure were initiated in late 2023. Expanding the scope to include more comprehensive and newer benchmarks, such as SWE-bench  dataset \cite{jimenez2023swe}, is a clear future work option to provide broader insights into code generation SLMs serving configurations.

Lastly, the reliance on the Transformers library for implementation introduces some limitations in flexibility, as other libraries might yield different performance results. 
For instance, ONNX’s used version still lacks kernels for operators in CUDA execution provider and falls back to the CPU execution provider in the version we used \cite{onnxruntime2024pythonapi}, causing data transfers and addition of new nodes to the executed model. This imbalance can overstate resource utilization advantages for $\langle TORCH, CUDA \rangle$ within CUDA configurations.
Future work could explore using alternatives to assess generalizability across different implementations.

\textbf{Conclusion validity.}
The statistical power of our experiments was carefully designed to ensure the reliability of our findings. To address this, we employed statistical methods, adhered closely to a methodically planned experimental procedure as detailed in Section~\ref{sec:research_methodology}, and ensured reproducibility through comprehensive documentation of our methods and data in the replication package. Considering various factors, treatments, and runs, we utilized a total of 600 (12 code generation SLMs, 5 deep learning serving configurations, 10 runs) samples to respond to our research questions, each sample representing a set of 164 inference requests. This approach, coupled with the execution of ten experimental runs and ensuring minimal background software interference, aims to fortify the reliability of our results. The procedure for replicating our work is thoroughly described, with datasets and the replication package openly accessible for validation.

\textbf{Construct validity.} To mitigate mono-method bias, we utilized two different profiling tools and a wattmeter to cross-verify the energy consumption recorded by the CPU (via energibridge) and GPU (via nvidia-smi) against the total energy reported by the wattmeter, ensuring the validity of our energy measurements. Additionally, we included twelve state-of-the-art SLMs in our study. Concerning model correctness, we did not evaluate this aspect as we employed well-established, previously validated models. All experimental outputs are provided in the replication package for comprehensive examination.

\textbf{Internal validity.} Potential internal validity threats such as hardware condition changes over successive experimental runs were addressed by implementing a five-minute period between each configuration and experiment run. This protocol helps maintain consistent hardware conditions across all tests. To minimize risks associated with the implementation of the inference process, we used default configurations provided by HuggingFace. While the repetition strategy and idle-time control minimized variability and helped ensure fair comparisons, fixed-order execution may still introduce minor risks related to system temperature changes or background activity fluctuations. Future studies could include shuffling of configurations to further mitigate such risks.

\textbf{External validity.} Our study's external validity is potentially limited by the specific subset of deep learning serving configurations and models used. The models and configurations were selected based on their current relevance and availability, but they may not encompass all possible variations in deep learning serving scenarios. To mitigate this, we incorporated the two most commonly used execution providers available to us and four well-known runtime engines. Although JIT was part of the initial design, it was excluded from evaluation due to the error-prone inference results.

Moreover, external validity is influenced by the representativeness of the used dataset itself. HumanEval contains basic to intermediate coding challenges, and our input dataset of 10 to 15 tokens captures only the short prompt auto-complete code task. Consequently, our findings are suited to interactive auto-complete scenarios and may not extend to generation that requires longer context, the synthesis of complete functions, or multi-file conversations. Future work should therefore adopt larger prompts and benchmarks that span multiple files to measure serving resource utilization under those conditions.

\section{Conclusions and future work}\label{sec:conclusions}

As outlined in the European Union AI Act \cite{euai2024act}, energy consumption is a critical factor in the lifecycle of deep learning models, particularly in both the training and inference phases. The Act emphasizes that energy efficiency must be considered as a key criterion when utilizing deep learning technologies. With GPUs now emerging as the most widely used hardware for training and deploying deep learning models due to their parallel processing capabilities, optimizing energy consumption has become even more pertinent.

The role of infrastructure-aware evaluation is expected to become increasingly important. For high-risk systems (as classified in the EU AI Act), the legislation introduces provisions that may require providers to meet energy-efficiency standards and to document the energy consumption of AI systems throughout their lifecycle, including during the inference phase (Art. 40) \cite{euai2024act}. These obligations are likely to encourage more detailed and standardized reporting practices. 
Future research can support this shift by developing methodologies to isolate not only the energy impact but also the resource utilization impact of serving infrastructure components, to clarify the distribution of resource consumption across lifecycle phases (e.g., training versus inference), and to analyze the software stack involved in inference. Figure~\ref{fig:adds_re_ep} provides a general visualization of the deep learning serving stack, which includes components such as runtime engines and execution providers, highlighted in red. These efforts can also contribute to the definition of benchmarks aligned with regulatory expectations. This regulatory direction may foster the adoption of more transparent, auditable, and resource-efficient deployment practices, creating new opportunities to enhance the sustainability of deep learning systems at scale.

In this work, we studied deep learning serving configurations, as duplets consisting of a runtime engine and an execution provider. We evaluated the energy consumption, execution time, and computing-resource utilization across five distinct configurations using twelve different code generation SLMs. Our observations reveal that:
\begin{enumerate}
    \item  The choice of runtime engine and execution provider significantly impacts energy consumption, execution time, and computing-resource utilization during deep learning inference.
    \item  Configurations utilizing the CUDA execution provider consistently demonstrate superior energy and time efficiency.
    \item  JIT engine shows promise as an optimal choice for specific scenarios, though it may require further enhancements to fulfill its potential for code generation.
    \item  In our experiments, the baseline TORCH engine coupled with the CUDA execution provider exhibited efficient energy consumption, execution time, and computing-resource utilization.
\end{enumerate}

The observed dominance of the $\langle TORCH, CUDA \rangle$ configuration reflects its maturity, extensive support, and integration within the PyTorch and HuggingFace ecosystem, rather than inherent model-specific optimization. Nonetheless, optimization effectiveness might vary for other models and tasks. 
To support serving infrastructure decisions in real-world tasks, we provide context-aware guidelines that help practitioners prioritize among energy consumption, execution time, and computing-resource utilization depending on system-level constraints.
These guidelines serve as a starting point for deploying similar decoder-based SLMs in various scenarios. For example, execution time may be prioritized in latency-critical settings such as IDE-based auto-complete code task, while energy efficiency becomes more relevant in battery-constrained or cost-sensitive deployments. In computing-resource utilization limited systems, minimizing RAM or VRAM usage may justify sacrificing some speed or energy performance. By framing serving infrastructure selection as a trade-off decision, our guidance enables engineers to tailor serving configurations to operational goals.  
We anticipate that future advancements in optimized runtime engines will enhance these metrics and address existing limitations, including error-proneness, to make them even more robust and efficient.
Consequently, our study should be viewed as a best-case snapshot for $\langle TORCH, CUDA \rangle$ on used hardware. As kernel coverage converges and libraries configurations such as IOBinding are tuned, alternative engines are likely to approach, or even surpass, this baseline \cite{stiles2024efficient}.

Our research highlights distinct resource utilization advantages among the various configurations. Specifically, the CUDA execution provider paired with TORCH consistently yielded the best overall resource utilization. Conversely, when utilizing the CPU execution provider, both OV and ONNX runtime engines enhanced resource utilization metrics significantly. In configurations with the CUDA execution provider and TORCH, energy consumption using this configuration can be from 10.84\% up to 62.01\%, and execution times from 10.26\% up to 52.16\% comparing to worst CUDA execution provider runtime engine (ONNX runtime). For setups using the CPU execution provider with OV or ONNX, we observed the energy consumption can be from 27.96\% up to 91.02\%, and execution times from 28.79\% up to 79.04\% compared to the baseline TORCH engine. Notably, the ONNX runtime engine did not exhibit superior resource utilization compared to the baseline TORCH configuration within CUDA execution provider configurations.

While JIT can offer performance or resource utilization benefits, it struggles with code generation. Future efforts might investigate alternative just-in-time approaches capable of handling branching logic more effectively. Overcoming these limitations would help to serve a broader range of use cases involving diverse inputs and execution paths.

Our findings explicitly underscore the complex, nuanced relationships among energy consumption, execution time, and computing resource utilization in SLMs inference infrastructures. We highlighted clear scenarios where optimizing execution time did not necessarily optimize energy consumption, reinforcing the explicit need for direct energy monitoring in practical settings.
Subsequent studies should explicitly focus on investigating these trade-offs across broader deep-learning scenarios and model architectures, aiming to provide more generalized frameworks and actionable guidelines for infrastructure optimization.

This work reinforces the importance of treating infrastructure components as architectural elements in deep learning system design. We recommend integrating the resource utilization profiling of serving infrastructure directly into development workflows and MLOps pipelines, helping software teams continuously evaluate trade-offs between energy, time, and computing-resource utilization across system updates and deployments.


Future research should replicate our experiments across a broader spectrum of deep learning tasks and input lengths. Because the present study concentrated on short prompts that represent auto-complete code task, further work is needed to verify whether the configurations advantages we observed hold when generating large-context or conversational code generation. Benchmarks such as SWE bench \cite{jimenez2023swe} and realistic multiple file tasks can reveal how serving configurations scale in these richer settings. Beyond code generation, investigations in areas like summarization and text classification will confirm the generalizability of our findings and may uncover insights that are specific to particular tasks.
Additionally, we recommend investigating newly developed runtime engines and execution providers. As technology advances, these newer configurations are likely to enhance performance and resource utilization efficiency, further promoting sustainable deep learning practices.

\section*{Acknowledgment}
This paper has been partly funded by the GAISSA Spanish research project (ref. TED2021-130923B-I00; MCIN/AEI/10.13039/501100011033).

\clearpage

\bibliographystyle{elsarticle-num-names} 

\bibliography{references}

\end{document}